\documentclass[epj]{svjourmod}

\usepackage{amsmath,amssymb,amsfonts}
\usepackage{graphicx}
\usepackage{citesort}
\usepackage{array} 

\allowdisplaybreaks[1]

\numberwithin{equation}{section}

\def\beq{\begin{equation}}
\def\eeq{\end{equation}}

\newcommand{\Eg}{E}
\newcommand{\Q}{\mathcal{Q}}

\providecommand{\tn}{\textnormal}
\providecommand{\amp}{\mathcal{A}}
\providecommand{\order}{\mathcal{O}}

\providecommand{\me}{M_\eta^2}
\providecommand{\mpi}{M_\pi^2}
\providecommand{\mpn}{M_{\pi^0}^2}
\providecommand{\mk}{M_K^2}
\providecommand{\mkn}{M_{K^0}^2}

\providecommand{\gev}{\,\tn{GeV}}
\providecommand{\mev}{\,\tn{MeV}}
\providecommand{\ev}{\,\tn{eV}}

\DeclareMathOperator{\Li}{Li}
\DeclareMathOperator{\re}{Re}
\DeclareMathOperator{\im}{Im}
\DeclareMathOperator{\ndf}{ndf}

\renewcommand{\thefigure}{\thesection.\arabic{figure}}
\renewcommand{\thetable}{\thesection.\arabic{table}}

\begin{document}


\title{Electromagnetic corrections in \boldmath{$\eta\to3\pi$} decays}
\titlerunning{Electromagnetic corrections in $\eta\to3\pi$ decays}

\author{Christoph Ditsche\inst{1}, Bastian Kubis\inst{1}, Ulf-G. Mei{\ss}ner\inst{1,2}}

\institute{
   Helmholtz-Institut f\"ur Strahlen- und Kernphysik (Theorie)
   and 
   Bethe Center for Theoretical Physics,
   Universit\"at Bonn, D-53115 Bonn, Germany
\and
   Institut f\"ur Kernphysik (Theorie), Institute for Advanced Simulations, 
   and J\"ulich Center for Hadron Physics, Forschungszentrum J\"ulich, 
   D-52425  J\"ulich, Germany
}

\authorrunning{C. Ditsche, B. Kubis, and U.-G. Mei{\ss}ner}

\date{
}

\abstract{
We re-evaluate the electromagnetic corrections to $\eta \to 3\pi$ decays
at next-to-leading order in the chiral expansion, arguing that effects of order $e^2(m_u-m_d)$ 
disregarded so far are not negligible compared to other contributions of order $e^2$ times
a light-quark mass.
Despite the appearance of the Coulomb pole in $\eta \to \pi^+\pi^-\pi^0$ and cusps
in $\eta\to3\pi^0$, the overall corrections remain small.
}

\maketitle

\setcounter{figure}{0}
\setcounter{table}{0}
\section{Introduction}\label{sec:introduction}

The decay $\eta\to3\pi$ is particularly interesting because it is forbidden by isospin symmetry. 
While the $\eta$ meson has isospin $I=0$, three pions with zero angular momentum can only 
couple to $I=1$, and thus the decay can only happen via isospin breaking $\Delta I=1$ operators. 
In the Standard Model, there are two such sources of isospin violation, 
on the one hand strong interactions from
\begin{equation}
 \mathcal{H}_\tn{QCD}(x)=\frac{m_d-m_u}{2}(\bar{d}d-\bar{u}u)(x)\;,
\end{equation}
which are proportional to the light-quark mass difference $m_d-m_u$, 
and on the other hand electromagnetic interactions that are proportional to 
the electric charge squared from
\begin{equation}
 \mathcal{H}_\tn{QED}(x)=-\frac{e^2}{2}\int dy\; D^{\mu\nu}(x-y)T(j_\mu(x)j_\nu(y))\;,
\end{equation}
where $D^{\mu\nu}(x-y)$ is the photon propagator and $j_\mu(x)$ is the current density 
containing the charged fields of the theory. 
Sutherland and Bell showed by using soft-pion techniques that the electromagnetic contribution 
at tree level is much too small to account for the observed decay rate~\cite{sutherland,bell}. 
The $\eta$ decay is therefore very sensitive to $m_d-m_u$ and hence it potentially 
yields a particularly clean access to the determination of quark mass ratios.
The strong tree-level amplitude was subsequently studied using current algebra
and partially conserved axial-vector current (PCAC) techniques~\cite{cronin,osborn},
but the decay width turned out to be off from the experimental value by a factor of a few. 
Later PCAC and current algebra were generalized and cast in a modern form 
in the framework of chiral perturbation theory (ChPT)~\cite{weinbergchpt,glannphys,glchpt}.
Gasser and Leutwyler (GL) subsequently calculated the strong contribution at one-loop level~\cite{gleta}. 
They observed large unitarity corrections due to strong final-state interactions,
but their value still differs from experiment by a factor of 2. 
Corrections beyond one loop were studied using dispersive 
methods~\cite{kambor,anisovich}, unitarized ChPT~\cite{Beisert,borasoy}, 
and finally with a complete two-loop calculation~\cite{bijnens}.
All of these find considerable enhancement of the decay widths compared
to the one-loop calculation.

Baur, Kambor, and Wyler (BKW)~\cite{bkw} studied corrections to Sutherland's theorem
by evaluating the electromagnetic contributions in $\eta\to 3\pi$ at one-loop level,
using an extension of ChPT including virtual-photon effects~\cite{urech}, 
but they found them to be very small.
The motivation for reconsidering these electromagnetic corrections at one-loop chiral order 
in the present work hinges on the fact that Ref.~\cite{bkw} neglects terms proportional to $e^2(m_d-m_u)$
(such terms have recently been considered for $\eta\to 3\pi^0$ in Ref.~\cite{Deandrea}), 
arguing that these are of second order in isospin breaking and therefore expected to be suppressed even further.
However, by restricting oneself to terms of the form $e^2 \hat m$ 
(where $\hat m=(m_u+m_d)/2$) and $e^2 m_s$, one excludes some of the most obvious
electromagnetic effects: real- and virtual-photon contributions, as well as 
effects due to the charged-to-neutral pion mass difference 
(which is predominantly of electromagnetic origin), both of which scale as $e^2(m_d-m_u)$.
These mechanisms fundamentally affect the analytic structure of the amplitudes in question:
in the charged decay channel $\eta\to\pi^+\pi^-\pi^0$, there is a Coulomb pole at the 
boundary of the physical region (at the $\pi^+\pi^-$ threshold), 
while in the neutral decay channel $\eta\to3\pi^0$, the pion mass difference 
induces a cusp behavior at the $\pi^+\pi^-$ thresholds (compare e.g.\ Refs.~\cite{MMS,UGMOsaka}).
This cusp encodes information on $\pi\pi$ scattering
in principle in much the same way as the decay $K^+ \to \pi^0\pi^0\pi^+$,
which has been established as a new means for a precision determination
of the scattering lengths combination 
$a_0-a_2$~\cite{cabibbo,isidori,na48,gamiz,colangelocusp,bisseggercusp,bisseggerradcorr}.
In contrast, the corrections identified in Ref.~\cite{bkw} are all polynomials
(due to counterterms) or quasi-polynomials (due to kaon loop effects) inside the physical region.
Furthermore, a soft-pion theorem~\cite{sutherland} guarantees that at the soft-pion point,
these corrections are of order $e^2 \hat m$ only (and not $\order(e^2 m_s)$);  
hence the relative suppression of the neglected terms is of the order of 
$(m_d-m_u)/(m_d+m_u)\approx1/3$ and therefore not a priori small.

This work is organized as follows. 
In Sect.~\ref{sec:formalism} we give a short introduction into the formalism of 
chiral perturbation theory with virtual photons and mention some special topics 
that are relevant for our investigation. 
The decay amplitudes for both channels are derived and displayed in Sect.~\ref{sec:amplitudes}, 
and in Sect.~\ref{sec:results} our numerical results are presented. 
Finally, we conclude with a summary and an outlook.
Various technical issues are relegated to the Appendices.

\setcounter{figure}{0}
\setcounter{table}{0}
\section{Formalism}\label{sec:formalism}

A particularly successful approach to describe the interactions of hadrons at low energies is 
to construct an effective field theory which encodes the infrared behavior of QCD. 
Based upon the approximate chiral symmetry of $\mathcal{L}_\tn{QCD}$ 
one can identify the lightest particles in the spectrum with the pseudo-Goldstone bosons 
induced by spontaneous breaking of this symmetry, and their interactions can be written 
as an expansion in small momenta and light-quark masses. 
This effective theory of the strong interactions is called chiral perturbation theory 
(ChPT)~\cite{weinbergchpt,glannphys,glchpt}. 
It can be generalized to include electromagnetic effects systematically~\cite{urech}
(see however Ref.~\cite{BernQEDQCD}). 
There are several good introductions to ChPT; see for example Refs.~\cite{scherer02,meissnerchpt,kubischpt}.

\subsection{Lagrangians}

\begin{sloppypar}
The Goldstone boson fields are collected in the field $U$ according to
\begin{equation}
 U =\exp \frac{i\phi}{F_0} \,,~
\phi=\sqrt{2}\begin{pmatrix}
\frac{\pi_3}{\sqrt{2}}+\frac{\eta_8}{\sqrt{6}}&\pi^+&K^+\\
\pi^-&-\frac{\pi_3}{\sqrt{2}}+\frac{\eta_8}{\sqrt{6}}&K^0\\
K^-&\bar{K}^0&-\frac{2\eta_8}{\sqrt{6}}\end{pmatrix} \,.
\end{equation}
The effective Lagrangian for the mesonic sector invariant under chiral symmetry 
reads at leading order (LO) $\order(p^2)$
\begin{equation}\label{eqn:l2strong}
 \mathcal{L}_\tn{str}^{(2)}=\frac{F_0^2}{4}\langle D^\mu U^\dagger D_\mu U + \chi^\dagger U+U^\dagger\chi\rangle\;,
\end{equation}
where $\langle...\rangle$ denotes the trace in flavor space. 
The external (pseudo-) scalar sources are conventionally combined into
\begin{equation}
 \chi=2B_0(s+ip)\;,\qquad s=\mathcal{M}+...\;,
\end{equation}
incorporating the quark mass matrix $\mathcal{M} = \mathrm{diag}(m_u,m_d,m_s)$.
$F_0$ is the pion decay constant in the chiral limit, 
and $B_0$ is related to the chiral quark condensate.
\end{sloppypar}

The terms of the Lagrangian at next-to-leading order (NLO) $\order(p^4)$ needed in the following are given by
\begin{align}\label{eqn:l4strong}
 \mathcal{L}_\tn{str}^{(4)} &= \nonumber
  L_3\langle D^\mu U^\dagger D_\mu U D^\nu U^\dagger D_\nu U\rangle\\\nonumber
 &+L_4\langle D^\mu U^\dagger D_\mu U\rangle\langle\chi^\dagger U+U^\dagger\chi\rangle\\\nonumber
 &+L_5\langle D^\mu U^\dagger D_\mu U(\chi^\dagger U+U^\dagger\chi)\rangle\\\nonumber
 &+L_6\langle\chi^\dagger U+U^\dagger\chi\rangle^2+L_7\langle\chi^\dagger U-U^\dagger\chi\rangle^2\\
 &+L_8\langle\chi^\dagger U\chi^\dagger U+U^\dagger\chi U^\dagger\chi\rangle\;.
\end{align}
Loops with vertices from the LO effective Lagrangian $\mathcal{L}_\tn{str}^{(2)}$ generate 
infinities that can be absorbed by a renormalization of the introduced NLO low-energy constants (LECs). 
For this purpose one defines renormalized low-energy constants
\begin{equation}
 L_i=\Gamma_i\lambda+L_i^r(\mu)\;,
\end{equation}
where $\lambda$ contains a pole in $d=4$ space-time dimensions, 
and the coefficients $\Gamma_i$ are given in Ref.~\cite{glchpt}. 
The renormalized coefficients depend on the scale $\mu$ introduced by dimensional regularization. 
All physical observables are finite and scale independent, which
serves as a check in explicit calculations.

\begin{sloppypar}
Following Ref.~\cite{urech} to include electromagnetism in the framework of ChPT 
by adding virtual photons as additional dynamical degrees of freedom, 
the full NLO effective Lagrangian can be written as
\begin{equation}
 \mathcal{L}_\tn{eff}=\mathcal{L}_\tn{str}^{(2)}+\mathcal{L}_\tn{em}^{(2)}
+\mathcal{L}_\tn{str}^{(4)}+\mathcal{L}_\tn{em}^{(4)}+\order(p^6)\;.
\end{equation}
The new local electromagnetic interactions in the Lagrangians $\mathcal{L}_\tn{em}^{(2)}$, $\mathcal{L}_\tn{em}^{(4)}$
contain the quark charge matrix $\Q=e\,\mathrm{diag}(2,-1,-1)/3$.
At leading order the local interactions stem from the following additional effective Lagrangian
\begin{equation}
 \mathcal{L}_\tn{em}^{(2)}=-\frac{1}{4}F^{\mu\nu}F_{\mu\nu}+C\langle \Q U\Q U^\dagger\rangle\;,
\end{equation}
where $F_{\mu\nu}$ is the electromagnetic field strength tensor, and
we have omitted the gauge fixing term. The new low-energy constant $C$ determines the 
purely electromagnetic part of the masses of the charged mesons in the chiral limit. 
For convenience one often defines the dimensionless constant
\begin{equation}
 Z=\frac{C}{F_0^4}\;,
\end{equation}
which we will use in the following.
\end{sloppypar}

At next-to-leading order the terms of the additional Lagrangian relevant in the following read~\cite{urech}
\begin{align}
 \mathcal{L}_\tn{em}^{(4)}& = \nonumber
 F_0^2 \Bigl\{ \langle D^\mu U^\dagger D_\mu U\rangle \bigl[ K_1 \langle \Q^2\rangle
+K_2 \langle \Q U\Q U^\dagger\rangle \bigr]\\\nonumber
 &+K_3 \bigl(\langle D^\mu U^\dagger \Q U\rangle\langle D_\mu U^\dagger \Q U\rangle
\\\nonumber &\qquad
+\langle D^\mu U\Q U^\dagger\rangle\langle D_\mu U\Q U^\dagger\rangle\bigr)\\\nonumber
 &+K_4 \langle D^\mu U^\dagger \Q U\rangle\langle D_\mu U\Q U^\dagger\rangle\\\nonumber
 &+K_5 \langle \bigl( D^\mu U^\dagger D_\mu U+D^\mu U D_\mu U^\dagger\bigr)\Q^2\rangle\\\nonumber
 &+K_6 \langle D^\mu U^\dagger D_\mu U\Q U^\dagger \Q U
+D^\mu U D_\mu U^\dagger \Q U\Q U^\dagger\rangle\\\nonumber
 &+\langle\chi^\dagger U+U^\dagger\chi\rangle \bigl[ K_7 \langle \Q^2\rangle
+K_8 \langle \Q U\Q U^\dagger\rangle \bigr]\\\nonumber
 &+K_9 \langle \bigl(\chi^\dagger U+U^\dagger\chi+\chi U^\dagger+U\chi^\dagger\bigr)\Q^2\rangle\\\nonumber
 &+K_{10/11} \langle(\chi^\dagger U \pm U^\dagger\chi)\Q U^\dagger \Q U
\\ &\qquad\qquad
+(\chi U^\dagger \pm U\chi^\dagger)\Q U\Q U^\dagger\rangle \Bigr\} \;, \label{eqn:l4em}
\end{align}
where we have neglected inter alia
terms proportional to $e^4$ (i.e.\ of second order in electromagnetic isospin breaking)
that are disregarded throughout this work. 
The electromagnetic LECs are renormalized in analogy to the strong LECs by
\begin{equation}
 K_i=\Sigma_i\lambda+K_i^r(\mu)\;,
\end{equation}
with the coefficients $\Sigma_i$ given in Ref.~\cite{urech}.

\subsection{Meson masses and mixing at leading order}

\begin{sloppypar}
By expanding the leading-order effective Lagrangian 
$\mathcal{L}^{(2)}=\mathcal{L}_\tn{str}^{(2)}+\mathcal{L}_\tn{em}^{(2)}$ 
one derives the LO meson masses. 
The flavor-neutral states $\pi_3$ and $\eta_8$ are mixed due to a difference 
in the light-quark masses $m_u-m_d\neq0$ according to
\begin{equation}
 \frac{B_0}{2}\begin{pmatrix}\pi_3\\\eta_8\end{pmatrix}^T\begin{pmatrix}
m_u+m_d&\frac{1}{\sqrt{3}}(m_u-m_d)\\
\frac{1}{\sqrt{3}}(m_u-m_d)&\frac{1}{3}(m_u+m_d+4m_s)\end{pmatrix}
\begin{pmatrix}\pi_3\\\eta_8\end{pmatrix} \,,
\end{equation}
which can be diagonalized by the single rotation angle\footnote{Note that this description 
of the $\pi^0\eta$ mixing via a single rotation angle is only valid at lowest chiral order~\cite{glchpt}.
}
\begin{equation}\label{eqn:eps}
 \epsilon= \frac{1}{2}\arctan\biggl(\frac{\sqrt{3}}{2}\frac{m_d-m_u}{m_s-\hat{m}}\biggr)
 = \frac{\sqrt{3}}{4}\frac{m_d-m_u}{m_s-\hat{m}}+\order(\delta^3)\;,
\end{equation}
where for convenience and later use we defined the average light-quark mass $\hat{m}$ 
and the isospin breaking parameter $\delta$,
\begin{equation}
 \hat{m}=\frac{m_u+m_d}{2}\;,\qquad\delta=m_d-m_u\;,
\end{equation}
leading to the physical meson mass eigenstates
\begin{equation}\label{eqn:leadmix}
 \begin{pmatrix}\pi^0\\\eta\end{pmatrix}=
\begin{pmatrix}\cos\epsilon&\sin\epsilon\\-\sin\epsilon&\cos\epsilon\end{pmatrix}
\begin{pmatrix}\pi_3\\\eta_8\end{pmatrix}\;.
\end{equation}
Mixing of the octet $\eta_8$ and the singlet $\eta_0$ to the observed mass eigenstates $\eta$ and $\eta'$ 
is encoded in the strong NLO low-energy constant $L_7$~\cite{leumixing}; the heavy $\eta'$ is 
not a dynamical particle in this theory.
\end{sloppypar}

For later use we define the charged-to-neutral meson mass differences
\begin{align}\label{eqn:massdifferences}
 \Delta\mpi &= \mpi\,-\,\mpn =2F_0^2Ze^2+\order(\delta^2,p^4)\;,
 \\\nonumber
 \Delta\mk &= \mk-\mkn = 2F_0^2Ze^2-B_0\delta+\order(p^4)\;,
\end{align}
where $M_\pi$, $M_K$ denote the charged-pion and -kaon masses.
From~\eqref{eqn:massdifferences} it is obvious that 
the LO electromagnetic contributions to the charged-pion and -kaon masses obey
Dashen's theorem~\cite{dashen},
\begin{equation}
 (\Delta\mpi)_\tn{em} = (\Delta\mk)_\tn{em} + \order(e^2m_q)\;.
\end{equation}
The Gell-Mann--Okubo (GMO) relation~\cite{gellmann,okubo}
including isospin violation is given by
\begin{equation}\label{eqn:gmo}
 2\mk+2\mkn-2\mpi+\mpn=3\me+\order(\delta^2,p^4)\;,
\end{equation}
which is fulfilled in nature to a few percent accuracy. 
We define the GMO discrepancy according to
\begin{equation}\label{eqn:deltagmo}
 \Delta_\tn{GMO}=\frac{2\mk+2\mkn-2\mpi+\mpn-3\me}{\me-\mpn} ~.
\end{equation}
We will frequently use the leading-order relations 
\begin{align}\label{eqn:msmhat}
 B_0\hat{m} &= \frac{\mpn}{2}+\order(\delta^2,p^4)\;,\nonumber\\
 B_0m_s &= \frac{3\me-\mpn}{4}+\order(\delta^2,p^4)\;, 
\end{align}
as well as~\eqref{eqn:massdifferences}
to replace quark masses and $Z$ by physical meson masses.

\subsection{Quark mass ratios}

The results of the previous section allow to form quark mass ratios 
determined entirely in terms of observable meson masses~\cite{weinbergratios},
\begin{align}\label{eqn:weinbergratios}
 \frac{m_u}{m_d} &\approx
\frac{\mk-\mkn-\mpi+2\mpn}{\mkn-\mk+\mpi}
\approx0.55\;,\nonumber \\
 \frac{m_s}{m_d} &\approx
\frac{\mkn+\mk-\mpi}{\mkn-\mk+\mpi}
\approx20.1\;, 
\end{align}
which are, however, subject to substantial higher-order corrections.
The double ratio
\begin{equation}\label{eqn:doubleratio}
 Q^2=\frac{m_s^2-\hat{m}^2}{m_d^2-m_u^2} 
\end{equation}
nevertheless is particularly stable with respect to strong higher-order corrections~\cite{glchpt},
\begin{equation}
 Q^2=\frac{\mk}{\mpi}\frac{\mk-\mpi}{(\mkn-\mk)_\tn{str}}\bigl\{1+\order(m_q^2,\delta,e^2)\bigr\}\;.
\end{equation}
$Q$ is the major semi-axis of Leutwyler's ellipse~\cite{leuellipse} 
(neglecting a tiny term proportional to $(\hat{m}/m_s)^2$),
\begin{equation}
 \left(\frac{m_u}{m_d}\right)^2+\frac{1}{Q^2}\left(\frac{m_s}{m_d}\right)^2=1\;.
\end{equation}
Furthermore, at leading order $Q^2$ is also invariant under a shift in the quark masses 
of the form $m_u\to m_u+\alpha \,m_dm_s$ (+ cyclic)~\cite{kaplan}.

By using Dashen's theorem in order to correct for the electromagnetic mass contributions,
the double ratio $Q^2$ can be calculated at LO yielding a numerical value of $Q_D\approx24.2$, 
which corresponds to the LO quark mass ratios~\eqref{eqn:weinbergratios}. 
However, Dashen's theorem is subject to potentially large higher-order corrections and different models 
(see e.g.\ Refs.~\cite{bdashen,dhwdashen,budashen}) yield a range
\begin{equation}
 1\lesssim\frac{(\mk-\mkn)_\tn{em}}{(\mpi-\mpn)_\tn{em}}\lesssim2.5\;,
\end{equation}
which leads to a rather large uncertainty in the numerical value of $Q$ of $20.6\lesssim Q\lesssim24.2\approx Q_D$.

Due to Sutherland's theorem~\cite{sutherland}, the dependence of the decay $\eta\to3\pi$ 
on the light-quark mass difference $m_d-m_u$ is much less prone to be obscured by electromagnetic
effects.  As will be shown below, the decay amplitudes can be written 
in terms of $Q^2$, and hence an accurate study of this decay can lead to an independent
determination of the quark mass ratios.  The re-evaluation of electromagnetic corrections
in the present work allows for an increased precision in this determination.

\setcounter{figure}{0}
\setcounter{table}{0}
\section{The decay amplitudes}\label{sec:amplitudes}

In this section we explain how to compute the decay amplitudes for both the charged and the neutral channel 
of $\eta\to3\pi$ decays at next-to-leading chiral order $p^4$ considering isospin breaking up to and 
including order $e^2(m_d-m_u)$, 
\begin{align}
 \langle\pi^0\pi^+\pi^-\;\tn{out}|\eta\rangle &=
i(2\pi)^4\delta^4(P_\tn{out}^c-p_\eta)\mathcal{A}_c(s,t,u)\;,\nonumber
 \\
 \langle3\pi^0\;\tn{out}|\eta\rangle &=i(2\pi)^4\delta^4(P_\tn{out}^n-p_\eta)\mathcal{A}_n(s,t,u)\;.
\end{align}
More technical details of the calculation can be found in Ref.~\cite{mythesis}.

\subsection{Leading-order decay amplitudes}

At leading chiral order $p^2$ one only has to compute one tree graph each 
since the fields in the lowest-order Lagrangian are already diagonalized by use of the mixing angle $\epsilon$.

\begin{figure}
 \centering
 \includegraphics[width=0.45\linewidth]{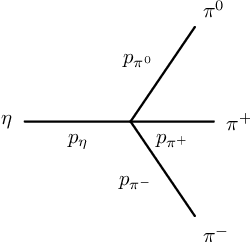} \hfill
 \includegraphics[width=0.45\linewidth]{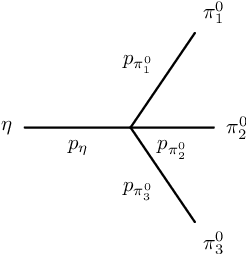}
 \caption{LO diagrams for the charged and neutral decays.}
 \label{fig:flo}
\end{figure}

The Feynman diagram describing the charged decay at lowest chiral order 
is shown in Fig.~\ref{fig:flo}.
The Mandelstam variables
\begin{equation}
 s=(p_\eta-p_{\pi^0})^2\;,\; t=(p_\eta-p_{\pi^+})^2\;,\; u=(p_\eta-p_{\pi^-})^2\;
\end{equation}
are related by
\begin{equation}
 s+t+u=\me+\mpn+2\mpi\equiv3s_0^c\;,
\end{equation}
where we defined the usual abbreviation $s_0^c$. 
Inserting the mixing angle $\epsilon$, 
expanding to leading order in $\delta$, and subsequently replacing $B_0m_s$, $B_0\hat{m}$, and $Z$ at 
leading chiral order by physical meson masses via~\eqref{eqn:msmhat} as well as $F_0$ by the 
pion decay constant $F_\pi$ yields the well-known lowest-order result for the amplitude,
\begin{align}\label{eqn:cloamp}
 \mathcal{A}_c^\tn{LO} &= -\frac{B_0(m_d-m_u)}{3\sqrt{3}F_\pi^2}\left\{1+
  \frac{3(s-s_0^c)+2\Delta\mpi}{\me-\mpn}\right\} \nonumber \\
  &= -\frac{(3s-4\mpn)(3\me+\mpn)}{Q^216\sqrt{3}F_\pi^2\mpn} \;.
\end{align}
An additional electromagnetic term of order $e^2(m_d-m_u)$ cancels the 
pion mass difference implicitly included in $s_0^c$.
The charged LO amplitude is completely proportional to $m_d-m_u \sim 1/Q^2$. 
It depends linearly on $s$, and by inserting $s_0^c$ it explicitly displays the Adler zero at $s=4\mpn/3$.

The neutral decay is described at leading order by the diagram shown in Fig.~\ref{fig:flo},
and in analogy to the charged decay we define the Mandelstam variables according to
\begin{equation}
 s=(p_\eta-p_{\pi^0_1})^2\;,\; t=(p_\eta-p_{\pi^0_2})^2\;,\; u=(p_\eta-p_{\pi^0_3})^2\;,
\end{equation}
which are related by
\begin{equation}\label{eqn:nsturelation}
 s+t+u=\me+3\mpn\equiv3s_0^n\;.
\end{equation}
The lowest-order amplitude contains neither derivatives nor electromagnetic terms 
and the consistent expansion in $\delta$ and the replacements described above yield
\begin{equation}\label{eqn:nloamp}
 \mathcal{A}_n^\tn{LO}=-\frac{3(\me-\mpn)(3\me+\mpn)}{Q^216\sqrt{3}F_\pi^2\mpn} \;.
\end{equation}
The neutral LO amplitude also has an overall factor of $m_d-m_u$, 
but is just a constant.

\subsection{Contributions at next-to-leading order}

\begin{sloppypar}
While for the charged process invariance under charge conjugation implies that the amplitude 
$\mathcal{A}_c(s,t,u)$ is symmetric under the exchange of $t$ and $u$,
\begin{equation}
 \mathcal{A}_c(s,t,u)=\mathcal{A}_c(s,u,t)\;,
\end{equation}
the amplitude $\mathcal{A}_n(s,t,u)$ for the neutral decay has to be symmetric under 
exchange of all pions and thus all Mandelstam variables due to Bose symmetry. 
All calculations of both $\eta\to 3\pi$ decay channels performed in ChPT so far~\cite{gleta,bkw,bijnens}
obey the following relation between the charged and the neutral amplitude
that utilizes isospin symmetry and the selection rule $\Delta I=1$,
\begin{equation}\label{eqn:isoamprel}
 \mathcal{A}_n(s,t,u)=\mathcal{A}_c(s,t,u)+\mathcal{A}_c(t,u,s)+\mathcal{A}_c(u,s,t)\;.
\end{equation}
For the LO amplitudes \eqref{eqn:cloamp} and \eqref{eqn:nloamp} 
this relation can be verified explicitly by use of \eqref{eqn:nsturelation}. 
However, the relation~\eqref{eqn:isoamprel} is only valid at leading order in isospin breaking 
and cannot be used at $\order(e^2(m_d-m_u))$. 
This is most easily seen by the fact that e.g.\ photon loop contributions
do not respect this relation.
Thus we have to calculate the neutral channel explicitly. 
\end{sloppypar}

\subsubsection{General procedure}

\begin{sloppypar}
In the following we will collect the contributing Feynman diagrams up to next-to-leading 
chiral order $p^4$ for both decays. In order to obtain an unambiguous and explicit ordering 
of the isospin breaking parameters $\{\delta,e^2\}$ with a minimal set of 
different meson masses we apply the following procedure.
\begin{enumerate}
 \item We rewrite the physical meson masses induced by derivatives in terms of 
$\me$, $\mpn$, $\delta$ and $\Delta\mpi\sim Ze^2$ by using the 
GMO relation~\eqref{eqn:gmo} and the charged-to-neutral mass differences~\eqref{eqn:massdifferences}.
Higher chiral orders can be neglected everywhere except for the lowest-order diagrams,
which we deal with explicitly in Sect.~\ref{sec:tree}.
 \item We insert the mixing angle $\epsilon$~\eqref{eqn:eps} and expand everything in terms of $\delta$.
 \item 
In order to disentangle the contributions to the various orders in isospin breaking
($\delta$, $e^2$, and $\delta e^2$), we expand the differences of certain
charged and neutral kaon loop functions, both for the tadpoles $\Delta_K-\Delta_{K^0}$ and for the two-point functions 
$J_{KK}-J_{K^0K^0}$ (see~\eqref{eqn:tadpole}, \eqref{eqn:loopJab1}, \eqref{eqn:loopJab2} in Appendix~\ref{app:mesonloops}).
We make use of the following expansion formulae, expressed in terms of $\delta$
and $\Delta\mpi=\order(e^2)$:
\begin{align}\label{eqn:kaontadexp}
 \Delta_K &-\Delta_{K^0}=(\Delta\mpi-B_0\delta)
\left[\frac{\Delta_{K^0}}{\mkn}+\frac{1}{16\pi^2}\right]\nonumber\\
 &-\frac{B_0\delta\Delta\mpi}{16\pi^2\mkn}+\order\bigl(e^4,\delta^2,p^4\bigr)\;, \\
\label{eqn:kaonloopexp}
 J_{KK}(s)& \!-\!J_{K^0K^0}(s)  =
\frac{2(B_0\delta\!-\!\Delta\mpi)}{s-4\mkn}\!\left[\!\bar{J}_{K^0K^0}(s)\!-\!\frac{1}{8\pi^2}\!\right]\nonumber\\
 &+\frac{4B_0\delta\Delta\mpi}{(s-4\mkn)^2}\bigg[
\bar{J}_{K^0K^0}(s)+\frac{s-8\mkn}{32\pi^2\mkn}
\bigg]\nonumber\\
 &+\order\bigl(e^4,\delta^2,p^2\bigr)\;.
\end{align}
We remark here in passing that kaon mass effects in loops as the above are the only source
of terms of $\order(\delta^2)$ in $\eta\to3\pi$ amplitudes. 
These are not of electromagnetic origin and therefore disregarded.
Their suppression with respect to the leading terms is never enhanced compared to
the numerical size of $\epsilon$ or $\Delta\mk/\mk$.
 \item We collect the terms with $\delta$, $e^2$, and $\delta e^2$ and neglect all terms proportional to $\delta^2$, $e^4$ 
and of higher isospin breaking orders.
 \item We factorize $1/Q^2\sim\delta$ where possible, using
\begin{equation}
 B_0\delta = \frac{3(\me-\mpn)(3\me+\mpn)}{Q^216\mpn} + \order(p^4) ~.
\end{equation}
 \item Finally, we replace left over quark masses and factors of $Z$ by the LO meson masses 
$\me$, $\mpn$, and $\Delta\mpi$ via~\eqref{eqn:msmhat} 
and~\eqref{eqn:massdifferences}, and $F_0$ by $F_\pi$.
\end{enumerate}
This scheme yields a representation in terms of the isospin breaking parameters $\{1/Q^2,e^2,\Delta\mpi\}$ 
and the meson masses $\me$ and $\mpn$. 
The latter choice is motivated by the fact the these masses appear as the asymptotic
states in the processes under investigation, and feature naturally at next-to-leading 
order in $\pi^0 \eta$ mixing, see Sect.~\ref{sec:tree}.
Note however that no mass substitutions take place inside loop functions, such that all cuts, 
even outside the physical region, remain at their exact places.
\end{sloppypar}

\subsubsection{Tree diagrams and mixing}\label{sec:tree}

The formalism to calculate matrix elements in the presence of mixing, using 
the Leh\-mann--Symanzik--Zimmer\-mann reduction formula, is explained in detail
in Refs.~\cite{abt,bijnens}.  Here we use an equivalent way to treat mixing, 
valid at one-loop order in ChPT, based on performing the calculation in a basis
of states diagonalized at tree level.

\begin{figure}
\centering
 \includegraphics[width=0.48\linewidth]{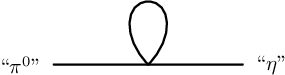} \hfill
 \includegraphics[width=0.48\linewidth]{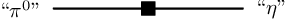}
 \caption{Diagrams contributing to $\pi^0\eta$ mixing at NLO.
 The quotation marks serve as a reminder that the $\pi^0$ and the $\eta$ cannot be on-shell at the same time.}
 \label{fig:fnlomix}
\end{figure}

Working in terms of tree-level eigenstates,
one has to account for $\pi^0\eta$ mixing effects beyond the leading
order relation~\eqref{eqn:eps}, \eqref{eqn:leadmix} at chiral order $p^4$, 
induced by the mixing diagrams Fig.~\ref{fig:fnlomix}.
Calculating the $\pi^0\eta$ self-energy matrix in the basis of tree-level eigenstates 
to one loop yields the off-diagonal element 
\begin{equation}
 \Sigma_{\pi^0\eta}(p^2)=Y_{\pi^0\eta}+Z_{\pi^0\eta}p^2\;,\quad Z_{\pi^0\eta}=\frac{\partial\Sigma_{\pi^0\eta}(p^2)}{\partial p^2}\;,
\end{equation}
written in a fashion that $Z_{\pi^0\eta}$ can be interpreted as the off-diagonal analogy to the 
wave-function renormalization factors.
E.g.\ in the charged decay channel, NLO mixing in the outgoing $\pi^0$ leg can be written in terms of
the LO amplitude $\amp_{\eta\to\eta\pi^+\pi^-}^\tn{LO}$ with subsequent mixing of the 
outgoing  $\pi^0$ due to the diagrams of Fig.~\ref{fig:fnlomix},
\begin{align}\label{eqn:mixdef}
 & \amp_{\eta\to\eta\pi^+\pi^-}^\tn{LO}  \!\times\! \frac{\Sigma_{\pi^0\eta}(\mpn)}{\mpn-\me} 
 \,=\, \amp_{\eta\to\eta\pi^+\pi^-}^\tn{LO}\biggl\{\frac{Z_{\pi^0\eta}}{2}+ \epsilon_4 \biggr\} \;, \nonumber
 \\
 & \epsilon_4 \,=\, \frac{1}{\mpn-\me}\,\Sigma_{\pi^0\eta}
\biggl(\frac{\mpn+\me}{2}\biggr)  \;
\end{align}
(see e.g.\ Ref.~\cite{ecker}).
In analogy the mixing of the ingoing $\eta$ with a $\pi^0$ is given by
\begin{equation}
 \amp_{\pi^0\to\pi^+\pi^-\pi^0}^\tn{LO} \!\times\! \frac{\Sigma_{\pi^0\eta}(\me)}{\me-\mpn} 
 \,=\, \amp_{\pi^0\to\pi^+\pi^-\pi^0}^\tn{LO}\biggl\{\frac{Z_{\pi^0\eta}}{2}-\epsilon_4\biggr\} \;.
\end{equation}
Since both $Z_{\pi^0\eta}$ and $\epsilon_4$ are of chiral order $p^2$ and of first order in isospin breaking, 
all quark mass replacements in the corresponding amplitudes can be done at order $p^2$ neglecting terms 
proportional to $\delta e^2$. The expressions for the corresponding LO amplitudes as well as
$Z_{\pi^0\eta}$ and~$\epsilon_4$ are displayed in Appendix~\ref{app:renorm}.
Taking into account the mixing of the ingoing $\eta$ and all outgoing $\pi^0$, 
the NLO mixing contributions for the 
charged and neutral channel can be summarized in the form
\begin{align}
 \amp_c^\tn{mix} &= \frac{3s-\me-\mpn}{3F_\pi^2}\;\biggl(\frac{Z_{\pi^0\eta}}{2} - \epsilon_4\biggr)
 +\frac{2\mpn}{3F_\pi^2}\;\epsilon_4 \;,\nonumber
 \\
 \amp_n^\tn{mix} &= \frac{\mpn}{F_\pi^2}\;Z_{\pi^0\eta} \;.
\end{align}

\begin{figure}
 \centering
 \includegraphics[scale=0.8]{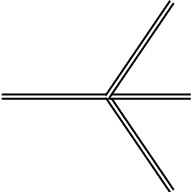} \hspace{12mm}  \includegraphics[scale=0.8]{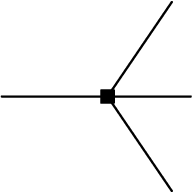}\\
 \hspace{7mm} a) Tree diagram \hspace{9mm} b) Counterterm diagram
 \caption{NLO tree-type diagrams.  The double solid line denotes the full propagator, cf.\ Fig.~\ref{fig:ffullprop}.}
 \label{fig:ftreetype}
\end{figure}

\begin{figure*}
 \centering
 \includegraphics[width=0.9\linewidth]{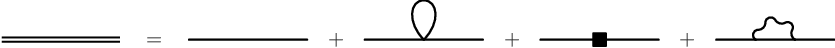}
 \caption{Full propagator up to $\order(p^4)$.  Wiggly lines denote photons.}
 \label{fig:ffullprop}
\end{figure*}

\begin{sloppypar}
With the mixing phenomenon dealt with, 
we can continue by considering the tree diagram displayed in Fig.~\ref{fig:ftreetype}a) 
consisting of a vertex of order $p^2$ and full propagators up to order $p^4$. 
The NLO propagator that is depicted in Fig.~\ref{fig:ffullprop} contains the LO propagator and 
proper self-energy insertions $-i\Sigma(p^2)$ of order~$p^4$ from meson tadpoles, 
from a  graph with a vertex from $\mathcal{L}^{(4)}$, 
and from photon loops for the charged pions.
The self-energies at chiral order $p^4$ allow to compute both the masses at NLO 
needed for renormalization corrections to the substitutions of quark masses by meson masses,
and the wave-function renormalization factors
\begin{equation}
 Z_a=\biggl\{1-\frac{\partial\Sigma_a(p^2)}{\partial p^2}\Big|_{p^2=M_a^2}\biggr\}^{-1}\equiv1+\Delta Z_a+\order(p^4)\;.
\end{equation}
Specifically, $\Sigma_{\pi^0}$ and $\Sigma_\eta$ refer to the diagonal elements
of the self-energy matrix in tree-level eigenstates.
By defining the following renormalization corrections
\begin{align}\label{eqn:rencorr}
 \me-\mpn &= \frac{4B_0}{3}(m_s-\hat{m})+\Delta_M\;,\nonumber
 \\
 \Delta\mpi &= 2F_0^2Ze^2+\Delta_{Z}\;,\nonumber
 \\
 \frac{3\me+\mpn}{4\mpn} &= \frac{m_s+\hat{m}}{2\hat{m}}\{1+\Delta_{Q}\}\;,\nonumber
 \\
 F_\pi &= F_0\{1+\Delta_{F_\pi}\}\;,
\end{align}
the corresponding NLO renormalization corrections to the LO amplitudes read
\begin{align}
 \amp_c^\tn{ren} &= -\frac{3\me+\mpn}{Q^216\sqrt{3}F_\pi^2\mpn}\Bigl\{(3s-4\mpn)(2\Delta_{F_\pi}-\Delta_Q)
\nonumber\\
&\qquad-\Delta_M-2\Delta_Z\Bigr\}  \;,\nonumber\\
 \amp_n^\tn{ren} &= -\frac{\sqrt{3}(3\me+\mpn)}{Q^216F_\pi^2\mpn}\Bigl\{(\me-\mpn)(2\Delta_{F_\pi}-\Delta_Q)
\nonumber\\
&\qquad-\Delta_M\Bigr\} \;.
\end{align}
The formulae for the $Z$ factors and the renormalization corrections defined above are shown
explicitly in Appendix~\ref{app:renorm}. 
Since the value for the physical pion decay constant
$F_\pi=92.2\,\tn{MeV}$~\cite{pdg} was extracted from decays of charged pions with 
electromagnetic corrections already taken care of (see also Ref.~\cite{holstein}),
we use the charged-pion decay constant in the absence of electromagnetism. 
The relation to the bare Goldstone boson decay constant $F_0$ is hence as given in Ref.~\cite{glchpt}.
\end{sloppypar}

Besides the renormalization effects discussed above, both  strong and  
electromagnetic NLO low-energy constants enter the calculation 
via the diagram Fig.~\ref{fig:ftreetype}b) with a vertex from 
$\mathcal{L}^{(4)}=\mathcal{L}_\tn{str}^{(4)}+\mathcal{L}_\tn{em}^{(4)}$. 
We denote the resulting contributions by 
$\amp_c^\tn{ct}$ and $\amp_n^\tn{ct}$, respectively.
For the charged channel one finds a contribution proportional to $L_3$ quadratic
in the Mandelstam variables, terms linear in $s$, and constant terms. 
However, the contribution for the neutral channel does not depend on $L_3$, $L_4$, 
and $K_1$ to $K_6$ (i.e.\ terms with traces containing only derivatives). 
Therefore it is a constant like the LO amplitude as well as other tree contributions.

Combining all parts discussed in this section, 
the tree diagram contributions up to next-to-leading chiral order then read
\begin{align}
 \amp_c^\tn{tree}&=\left\{1+\frac{\Delta Z_\eta}{2}+\frac{\Delta Z_{\pi^0}}{2} + \Delta Z_\pi \right\}\mathcal{A}_c^\tn{LO}
 \nonumber \\
& \qquad +\amp_c^\tn{ren} +\amp_c^\tn{mix} + \amp_c^\tn{ct} ~,\nonumber
 \\
 \amp_n^\tn{tree}&=\left\{1+\frac{\Delta Z_\eta}{2}+\frac{3\Delta Z_{\pi^0}}{2}\right\}\mathcal{A}_n^\tn{LO}
\nonumber\\ &\qquad  
+\amp_n^\tn{ren}+\amp_n^\tn{mix} + \amp_n^\tn{ct} ~.
\end{align}
In $\amp_{c/n}^\tn{tree}$, any explicit dependence on $L_4$ and $L_6$ cancels.
The remaining strong LECs are replaced in terms of observables in the spirit 
of Ref.~\cite{gleta}. 
Besides the correction $\Delta_\tn{GMO}$ to the Gell-Mann--Okubo relation~\eqref{eqn:deltagmo}, 
we define the correction $\Delta_F$ to the ratio of the kaon and the pion decay constants 
at NLO without isospin breaking as given in Ref.~\cite{glchpt} according to
\begin{equation}\label{eqn:deltakpi}
 \frac{F_K}{F_\pi}=1+\Delta_F ~.
\end{equation}
The explicit forms of these two quantities are given in Appendix~\ref{app:renorm}.
Since both $\amp_{c/n}^\tn{tree}$ depend only on the combination $2L_7+L_8$, both
$L_7$ and $L_8$ can be eliminated simultaneously in terms of $\Delta_\tn{GMO}$, 
while $L_5$ can be replaced by $\Delta_F$. 
The explicit formulae for $\amp_c^\tn{tree}$ and $\amp_n^\tn{tree}$ following from these replacements are 
given in Appendices~\ref{app:charged}, \ref{app:neutral}. 
In this form, the neutral amplitude has no explicit dependence on any strong LEC,
while the charged one only contains a term proportional to $L_3$.
Furthermore, as a by-product, in both any explicit dependence on the electromagnetic
constants $K_7$ and $K_8$ cancels.

\subsubsection{Meson loops}

\begin{figure}
 \centering
 \includegraphics[scale=0.8]{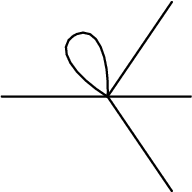} \hspace{10mm} \includegraphics[scale=0.8]{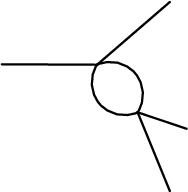}\\
 \hspace{5mm} a) Tadpole diagrams \hspace{5mm} b) Rescattering diagrams
 \caption{Meson loop contributions.}
 \label{fig:fmesonloops}
\end{figure}

At chiral order $p^4$ hadronic loops can occur in three places. 
In addition to the tadpole contributions to the NLO propagator that have been discussed already, 
there are also diagrams with a meson tadpole directly at the LO vertex and 
diagrams with rescattering intermediate mesons as depicted in Fig.~\ref{fig:fmesonloops}.

The results for the tadpole contributions $\amp_c^\tn{tad}$ and $\amp_n^\tn{tad}$ are given 
in Appendices~\ref{app:charged}, \ref{app:neutral}. 
Again the charged contribution contains terms linear in $s$ and constant terms, 
whereas the neutral contribution is constant.

The rescattering of intermediate mesons, in particular $I=0$ $\pi\pi$ final-state interaction,
is very important and gives rise to about half of the total NLO corrections for the decay $\eta\to3\pi$~\cite{gleta}. 
As pointed out before the charged decay is symmetric under $t\leftrightarrow u$ and the neutral decay is 
symmetric under exchange of all Mandelstam variables. 
While for the neutral decay it turns out that the contributions in each channel depend on the corresponding 
variable only, in case of the charged decay both the $t$ and the $u$ channel contributions depend on 
at least two variables. 
Since there are no virtual-photon loops for the neutral decay, 
these are the only dependences on Mandelstam variables. 

Defining the functions $\amp_c^{s}$, $\amp_c^{tu}$ and $\amp_n^{stu}$ as explained in Appendices~\ref{app:charged},
\ref{app:neutral}, 
the rescattering contributions for both decays read
\begin{align}\label{eqn:ampscatt}
 \amp_c^\tn{scatt}&=\amp_c^{s}(s)+\amp_c^{tu}(s,t)+\amp_c^{tu}(s,u)\;,\nonumber
 \\
 \amp_n^\tn{scatt}&= \amp_n^{stu}(s)+\amp_n^{stu}(t)+\amp_n^{stu}(u) \;.
\end{align}

\subsubsection{Photon loops}

\begin{figure}
 \centering
 \hspace{2mm} \includegraphics[scale=0.8]{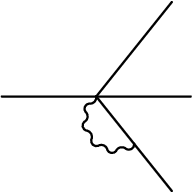} \hspace{12mm} \includegraphics[scale=0.8]{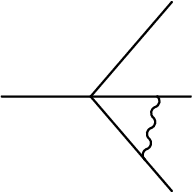}\\
 a) Vertex correction diagrams \hspace{2mm} b) Triangle diagram
 \caption{Virtual-photon loop contributions.}
 \label{fig:fphotonloops}
\end{figure}

For the charged decay at chiral order $p^4$, photon loops occur. 
These typically entail infrared (IR) divergences, which are cancelled in the standard manner
by the inclusion of real-photon radiation (bremsstrahlung) on the cross-section level.
We keep track of the infrared divergences by introducing a small photon mass $m_\gamma$;
alternatively one might use dimensional regularization also in the infrared.

In addition to the self-energy contributions to the NLO propagators of the charged pions, 
there are also vertex corrections as depicted in Fig.~\ref{fig:fphotonloops}a), 
and the triangle diagram displayed in Fig.~\ref{fig:fphotonloops}b), 
describing the exchange of a virtual photon in the final state between the charged pions.

The photon loop contribution $\Sigma_\pi^\gamma(p^2)$ to the self-energy of the charged pions 
contributes both to the (IR-finite) mass and the (IR-divergent)
wave-function renormalization. 
The two vertex correction diagrams turn out to yield the same IR-finite contribution $\amp_c^{\pi\gamma}$ 
and the triangle diagram leads to the contribution $\amp_c^{\pi\pi\gamma}$. 
Both are given explicitly in Appendix~\ref{app:charged}. The latter one contains the triangle loop function 
$G(s)$~\eqref{eqn:triangleloop} which has some interesting features: 
both the real and the imaginary part are IR-divergent; 
while the infrared divergence in the real part is cancelled against bremsstrahlung contributions, 
the imaginary part can be resummed in the (divergent) Coulomb phase.
Furthermore $G(s)$ contains a kinematical singularity at threshold $s=4\mpi$, the Coulomb pole. 
Its contribution to the complete amplitude is given by
\beq
\amp_c^\tn{pole} = \amp_c^\tn{LO} \times e^2 \frac{1+\sigma^2}{16\sigma} ~,
\label{eqn:CoulombPole}
\eeq
where $\sigma=\sqrt{1-4\mpi/s}$. 
Note that the prefactor of the triangle loop function, i.e.\ the Coulomb pole, 
is proportional to the LO amplitude~\eqref{eqn:cloamp}.

\subsubsection{Real-photon radiation}\label{sec:brems}

\begin{figure}
 \centering
 \includegraphics[scale=0.8]{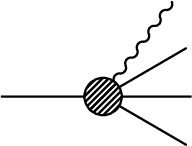}
 \caption{Real-photon radiation diagram.}
 \label{fig:fbrems}
\end{figure}

Infrared divergences due to virtual-photon corrections are canceled by including real-photon 
radiation; see Fig.~\ref{fig:fbrems}.
We only include real-photon radiation in the soft-photon approximation
that amounts to neglecting the photon momentum in the overall energy and momentum conservation.
The results for general $n$-body decays 
with one additional real photon of maximum energy $\Eg_\tn{max}$ 
radiated can be found, e.g., in Ref.~\cite{isidorisoftphoton}
(also using a finite photon mass as a regulator); radiative corrections for the largely analogous
decay $K_L \to \pi^+ \pi^- \pi^0$ in the framework of ChPT are discussed in Ref.~\cite{bijnensborg}.
Adapting the results of Ref.~\cite{isidorisoftphoton} to
$\eta\to\pi^+\pi^-\pi^0\gamma$ yields an effective contribution to the amplitude squared of the form
\begin{equation}\label{eqn:bremsdivparts}
 |\amp_c|^2\frac{e^2}{4\pi^2}\biggl[
 \ln\frac{m_\gamma^2}{4\Eg_\tn{max}^2}\left\{1-\frac{1+\sigma^2}{2\sigma}\ln\frac{1+\sigma}{1-\sigma}\right\}+F(s,t,u)\biggr]\;.
\end{equation}
The maximum photon energy $E_\tn{max}$ (in the $\eta$ rest frame) is given by
\begin{equation}\label{eqn:Emax}
 \Eg_\tn{max} = \min\{\Eg_\tn{kin},\Eg_\tn{cut}\} \;, \quad
 \Eg_\tn{kin} = \frac{\me-\left(M_{\pi^0}+\sqrt{s}\right)^2}{2M_\eta} \;,
\end{equation}
where $\Eg_\tn{kin}$ is the maximal kinematical limit, 
and the value of the photon cutoff energy $\Eg_\tn{cut}$ is set to a typical detector resolution. 
The kinematical constraint leads to a logarithmic divergence at the upper 
limit $s=(M_\eta-M_{\pi^0})^2$, signalling that very close to the boundary of phase space 
the $\order(\alpha)$ approximation becomes unreliable.
$F(s,t,u)$ represents contributions that are finite in the limit $m_\gamma \to 0$.
They are given explicitly as
\begin{align}
 F(s,t,u) &= f(1) + f(-1) - \int_{-1}^1 dz \frac{1+\sigma^2}{1-z^2\sigma^2} f(z) \;, \nonumber \\
 f(z) &= \frac{1+z\nu}{2\omega(z)}\ln\frac{1+z\nu+\omega(z)}{1+z\nu-\omega(z)} \;, \nonumber \\
 \omega(z) &= \sqrt{\beta^2+2z\nu+z^2(4\rho^2+\nu^2)} \;, \nonumber\\
 \beta &= \frac{\lambda^{1/2}(\me,s,\mpn)}{\me+s-\mpn} \;, \quad
 \nu = \frac{t-u}{\me+s-\mpn} \;, \nonumber \\
 \rho &= \frac{M_\eta\sqrt{s-4\mpi}}{\me+s-\mpn} \;,
\end{align}
where $\lambda(a,b,c)=a^2+b^2+c^2-2(ab+bc+ca)$ is the K\"{a}ll\'{e}n function.

Computing the amplitude for $\eta\to\pi^+\pi^-\pi^0$ squared and collecting the IR-divergent parts
consistently with respect to chiral order and power counting in  $e^2$ we obtain
\begin{equation}\label{eqn:irdivparts}
-|\mathcal{A}_c|^2\frac{e^2}{4\pi^2}\ln\frac{m_\gamma^2}{\mpi}\left\{1-\frac{1+\sigma^2}{2\sigma}
\ln\frac{1+\sigma}{1-\sigma}\right\} \;.
\end{equation}
Comparing~\eqref{eqn:bremsdivparts} and \eqref{eqn:irdivparts}
shows that the inclusion of bremsstrahlung cancels the IR-divergences. 

We have been intentionally vague above about the amplitude $\amp_c$ to be inserted in~\eqref{eqn:bremsdivparts}, 
\eqref{eqn:irdivparts}.  As a matter of principle, our calculation is only fully consistent 
for the lowest-order amplitude $\amp_c^\tn{LO}$, to be multiplied with electromagnetic corrections factors.
However, as is conventionally done in ChPT calculations, we only chirally expand the amplitudes to a specific
chiral order, and we do not re-expand their squares.  This way, even when disregarding terms of order $e^4$
(which are tiny), we include interference terms of radiative corrections with strong loop corrections.
As a consequence, in order to achieve cancellation of infrared divergences,
the bremsstrahlung terms in~\eqref{eqn:bremsdivparts} are included with a prefactor
\beq
|\amp_c|^2 \to \amp_c^\tn{LO}\bigl(\amp_c^\tn{LO}+\re\bigl\{\amp_c^\tn{NLO}|_{e=0}\bigr\}\bigr) ~.
\eeq

\subsubsection{Subtraction of universal corrections}\label{sec:universal}

\begin{sloppypar}
The kinematical singularities in the radiative corrections both at $s=4\mpi$~\eqref{eqn:CoulombPole}
and at $s=(M_\eta-M_{\pi^0})^2$~\eqref{eqn:bremsdivparts}, \eqref{eqn:Emax} are part of the 
universal soft-photon corrections that can even be resummed to all orders
in the fine structure constant~\cite{Yennie,WeinbergPhotons,isidorisoftphoton}.
In order to perform a meaningful fit of the Dalitz plot distribution (see Sect.~\ref{sec:Dalitz}),
these universal soft-photon corrections are usually already applied in the analysis of the 
experimental data.  We account for this by subtracting the corresponding parts of our amplitude (squared)
before predicting experimental observables, in the following way.
\begin{enumerate}
\item 
The both infrared and kinematically divergent imaginary part of the photon triangle loop function 
$G(s)$ \eqref{eqn:triangleloop} can be resummed to the (divergent) Coulomb phase.  As an overall
phase factor is unobservable, $\im \{G(s)\}$ is omitted.
\item
We subtract the Coulomb pole, whose leading approximation in our calculation is given in~\eqref{eqn:CoulombPole}.
\item
We subtract the $e^2 \ln E_\tn{max}$ singularity of~\eqref{eqn:bremsdivparts} in the form
\beq
 |\amp_c|^2\frac{e^2}{4\pi^2}
 \ln\frac{4\Eg_\tn{max}^2}{\me}\left\{\frac{1+\sigma^2}{2\sigma}\ln\frac{1+\sigma}{1-\sigma}-1\right\} ~,
\label{eqn:univLogE}
\eeq
which is the $\order(e^2)$ approximation of the resummed correction factor, 
see Refs.~\cite{WeinbergPhotons,isidorisoftphoton}.
\end{enumerate}
In this manner, we obtain a squared amplitude free of kinematical singularities.
\end{sloppypar}

For the purpose of illustration of the size of the various corrections, 
we find it useful to plot (real and imaginary parts of) \emph{amplitudes} in Sect.~\ref{sec:Namps},
which, as illustrated in the previous section, can in principle not be made infrared finite in a consistent way.  
We remedy this problem by hand, using the replacement 
\beq
m_\gamma \to M_\eta \label{eqn:IRcure}
\eeq 
in the decay amplitude, 
compare~\eqref{eqn:bremsdivparts}, \eqref{eqn:irdivparts}, \eqref{eqn:univLogE}.  
This does not take into account the finite contributions $F(s,t,u)$, 
which can only be added on the level of squared matrix elements.
Furthermore, in Sect.~\ref{sec:Namps} we retain the threshold divergences 
(Coulomb pole and phase) for illustrative reasons.

\subsection{Total decay amplitudes}

In terms of the contributions defined before and by applying the replacement rules given above the 
decay amplitudes finally can be written as
\begin{align}
 \mathcal{A}_c &= \amp_c^\tn{tree}+\amp_c^\tn{tad}+\amp_c^\tn{scatt}+2\amp_c^{\pi\gamma}+\amp_c^{\pi\pi\gamma}\;,\nonumber
 \\
 \mathcal{A}_n &= \amp_n^\tn{tree}+\amp_n^\tn{tad}+\amp_n^\tn{scatt}\;. \label{eqn:Atotal}
\end{align}
Note that in case of the neutral decay the whole kinematical dependence is contained in the 
scattering contribution, which takes the symmetric form~\eqref{eqn:ampscatt}.

\begin{sloppypar}
We have checked both the charged and the neutral amplitude in several ways. 
They are finite and renormalization-scale independent. 
In addition, we have explicitly verified that both amplitudes reduce to the results of Ref.~\cite{glchpt} 
at $\order(\delta)$ and of Ref.~\cite{bkw} at $\order(e^2)$.
\end{sloppypar}

We have pointed out before that the total amplitudes \eqref{eqn:Atotal} do not 
obey the $\Delta I=1$ selection rule~\eqref{eqn:isoamprel}.  
The terms of $\order(\delta e^2)$ violating this rule are, however, difficult
to write down or even define consistently, as the pion mass difference affects even 
the relation between the Mandelstam variables in the two channels (due to $s_0^c \neq s_0^n$).
Just for the purpose of illustration, we show here the part of the deviation
proportional to the various low-energy constants, using $s+t+u=3s_0^n$: 
\begin{align}
&\amp_n^\tn{LEC}-\amp_c^\tn{LEC}(s,t,u)-\amp_c^\tn{LEC}(t,u,s)-\amp_c^\tn{LEC}(u,s,t) \nonumber\\
&= \frac{3\me+\mpn}{\sqrt{3}Q^2 F_\pi^2 \mpn}\biggl\{ \frac{L_3}{F_\pi^2}\bigl(\me-3\mpn\bigr)\Delta\mpi \nonumber\\
&\quad + e^2 \biggl[ \frac{3}{8}(2K_3^r-K_4^r)\bigl(\me+\mpn\bigr) -K_6^r \me \nonumber\\
&\qquad  +\frac{3}{2}\bigl(K_{10}^r+K_{11}^r\bigr)\bigl(3\me-\mpn\bigr) \biggr] \biggr\} = \order(\delta e^2) \;.
\end{align}

\begin{sloppypar}
As a final analytic result, we want to comment on the soft-pion theorem~\cite{sutherland} 
for the electromagnetic corrections.  As shown in Ref.~\cite{bkw}, the corrections of $\order(e^2 m_q)$
(where $m_u=m_d$ is assumed)
at the kinematical point $s=3s_0^{c/n}$, $t=u=0$ (the soft-pion point) scale as $e^2 \hat m$, not as $e^2 m_s$.  
Explicitly, we find for the electromagnetic amplitudes in this approximation (denoted by BKW)
\begin{align}
\amp_c^\tn{BKW}&\stackrel{\tn{SP}}{=} \frac{\mpn}{\sqrt{3}F_\pi^2} \biggl\{ \frac{5\Delta\mpi}{16\pi^2F_\pi^2}
  \biggl(1-\sqrt{2}\arctan\frac{1}{\sqrt{2}}\biggr) \nonumber\\
& + \frac{4e^2}{3}\biggl[\frac{3}{2}(2K_3^r-K_4^r)-K_5^r-K_6^r+K_9^r+K_{10}^r \biggr]\biggr\} \nonumber \\
& + \order\Bigl(e^2 \frac{\hat m^2}{m_s}\Bigr) ~, \label{eqn:softCbkw}\\
\amp_n^\tn{BKW} &\stackrel{\tn{SP}}{=} \frac{\mpn}{\sqrt{3}F_\pi^2} \biggl\{ \frac{\Delta\mpi}{16\pi^2F_\pi^2}
  \biggl(1-\sqrt{2}\arctan\frac{1}{\sqrt{2}}\biggr) \nonumber\\
& + \frac{4e^2}{3}\biggl[\frac{3}{2}(2K_3^r-K_4^r)-K_5^r-K_6^r+K_9^r+K_{10}^r \biggr]\biggr\} \nonumber \\
& + \order\Bigl(e^2 \frac{\hat m^2}{m_s}\Bigr) ~, \label{eqn:softNbkw}
\end{align}
where $\stackrel{\tn{SP}}{=}$ denotes the evaluation at the soft-pion point
and the $K_i^r$ are to be evaluated at the scale $\mu=M_K$.
The additional terms of $\order(e^2\delta)$ (denoted by DKM), for comparison, are found to be
\begin{align}
\amp_c^\tn{DKM}&\stackrel{\tn{SP}}{=} -\frac{2B_0\,\delta}{\sqrt{3}F_\pi^2} \biggl\{ \frac{\Delta\mpi}{\me} 
\biggl[ 1+ \frac{4}{3}\Bigl( \Delta_\tn{GMO} + \Delta_F + \frac{\me}{F^2}L_3 \!\Bigr) \nonumber\\
&\qquad + \frac{\me}{16\pi^2F_\pi^2} \biggl( \frac{2}{3}\ln\frac{M_\pi}{M_K} - \frac{9}{2}\ln\frac{4}{3} + 5\nonumber\\
& \qquad \qquad \qquad
+\frac{7}{4\sqrt{2}}\arctan\frac{1}{\sqrt{2}} + 2\pi\,i\biggr)\biggr] \nonumber\\
& - \frac{e^2}{8\pi^2}\biggl[
 \Bigl(1+2\ln\frac{M_\pi}{M_\eta}+\pi\,i\Bigr)\ln\frac{m_\gamma}{M_\eta} - \ln^2\frac{M_\pi}{M_\eta} \nonumber\\
& \qquad - \frac{\pi^2}{3}+\frac{9}{4}+\frac{\pi}{4}\,i + \ln\frac{M_\pi}{M_K} +\frac{1}{4} \ln\frac{4}{3} \biggr] \nonumber\\
& - \frac{4e^2}{3}\biggl[K_1^r+K_2^r-\frac{7}{4}(2K_3^r-K_4^r)+K_5^r+K_6^r \nonumber\\
& \qquad -K_9^r+2K_{10}^r +3K_{11}^r \biggr]\biggr\} 
 + \order\Bigl(e^2 \frac{\delta\,\hat m}{m_s}\Bigr) , \hspace*{-1mm} \label{eqn:softCdkm}\\
\amp_n^\tn{DKM} &\stackrel{\tn{SP}}{=} -\frac{2B_0\,\delta}{\sqrt{3}F_\pi^2} \biggl\{ \frac{\Delta\mpi}{\me} \biggl[ \frac{8}{3}\Delta_F
+ \frac{\me}{16\pi^2F_\pi^2} \biggl( \ln\frac{M_\pi}{M_K} \nonumber\\
& \qquad + 2\ln\frac{4}{3} + 3
+\frac{1}{4\sqrt{2}}\arctan\frac{1}{\sqrt{2}} - \pi\,i\biggr)\biggr] \nonumber\\
& - \frac{4e^2}{3}\biggl[\frac{3}{16\pi^2} + K_1^r+K_2^r-\frac{5}{2}(2K_3^r-K_4^r)+K_5^r \nonumber\\
& \qquad +K_6^r-K_9^r+5K_{10}^r +6K_{11}^r \biggr]\biggr\} 
 + \order\Bigl(e^2 \frac{\delta\,\hat m}{m_s}\Bigr) \,, \label{eqn:softNdkm}
\end{align}
again at  $\mu=M_K$.
The comparison of ~\eqref{eqn:softCbkw}--\eqref{eqn:softNdkm} demonstrates explicitly
that terms of $\order(e^2\delta)$ are relatively suppressed only by $(m_d-m_u)/(m_d+m_u)\approx 1/3$
and not by another small isospin-violating parameter.  Furthermore, only the terms in~\eqref{eqn:softCdkm},
\eqref{eqn:softNdkm} show chiral logarithms $\propto \ln M_\pi$.  
The chiral logarithms squared in~\eqref{eqn:softCdkm} are due to the photon triangle loop function $G(s)$, 
expanded around the soft-pion point.
\end{sloppypar}

\setcounter{figure}{0}
\setcounter{table}{0}
\section{Numerical results}\label{sec:results}

The numerical input that was used to obtain the following results 
is collected in Appendix~\ref{app:numbers}.

\subsection{Kinematical bounds and cusps}

\begin{figure}
 \centering
 \includegraphics[width=\linewidth]{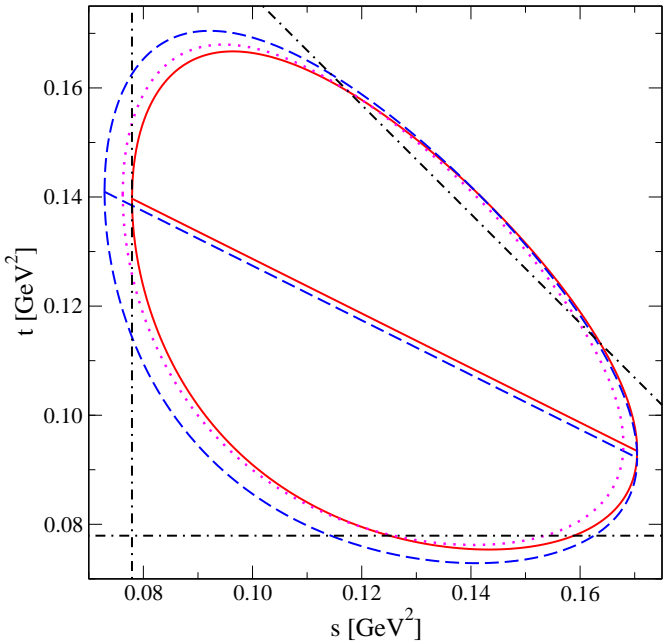} \vspace{-3mm}
\caption{Kinematical bounds of the Dalitz plots for the decays $\eta\to\pi^+\pi^-\pi^0$ (full/red), 
  $\eta\to3\pi^0$ (dashed/blue), and $\eta\to\pi^+\pi^-\pi^0$ with an average pion mass $\bar{M}_\pi^2$ 
  (dotted/magenta, see main text for details).
  Furthermore shown are the lines with $t=u$ for both the charged and the neutral decay 
  as well as the $4\mpi$ thresholds in all three kinematical channels (dashed-dotted/black).}
 \label{fig:dalitz}
\end{figure}

In Fig.~\ref{fig:dalitz} we plot the boundaries of the Dalitz plot for
both decay channels.
In addition to the charged and the neutral decay phase spaces we have also drawn 
the allowed phase space for the decay $\eta\to\pi^+\pi^-\pi^0$ using
an average pion mass $\bar{M}_\pi^2=(\mpn+2\mpi)/3$. 
This approximation is used frequently in the literature, since 
it is, strictly speaking, not possible to consistently account for different pion masses 
(which feature in $\eta\to3\pi$ at $\order(\delta e^2)$)
at leading order in isospin breaking.
Deviations at the border of the Dalitz plot are clearly visible.

\begin{sloppypar}
Recently, there has been a strong renewed interest in threshold cusp phenomena~\cite{wigner}
(for an overview of various examples, see also Ref.~\cite{Rosner})
since Cabibbo~\cite{cabibbo,isidori} pointed out that the large-statistics
experimental data on the decay $K^+\to\pi^0\pi^0\pi^+$ collected by the 
NA48/2 Collaboration at CERN may allow for a very precise determination
of the pion--pion scattering length combination $a_0-a_2$~\cite{na48}.  
These scattering lengths multiply the strength of the cusp in the 
$\pi^0\pi^0$ invariant mass distribution at the $\pi^+\pi^-$ threshold;
they are, at the same time, predicted theoretically with tremendous precision,
$a_0-a_2=0.265\pm0.004$~\cite{CGL,CGL2}, and represent a core test of
ChPT and our picture of chiral symmetry breaking.
The tailor-made theoretical framework to perform a precision analysis of these cusps
is non-relativistic effective field theory~\cite{colangelocusp,bisseggercusp,bisseggerradcorr}
that has been exploited up to two loops and including radiative corrections.
\end{sloppypar}

A similar cusp phenomenon also occurs in the decay $\eta\to3\pi^0$,
although, as in the analogous decay $K_L\to3\pi^0$~\cite{KTeV}, it is less pronounced 
than in $K^+\to\pi^0\pi^0\pi^+$.
From an experimental point of view $\eta\to3\pi^0$ decays have therefore 
not been competitive for a pion--pion scattering length determination. 
But as new experiments like WASA-at-COSY~\cite{wasa,WASAatCOSY} or  
Crystal Ball at MAMI-B~\cite{CBatMAMI,unverzagtPhD,unverzagt,prakhov}
will provide more high-statistics data in the near future, 
an investigation of the cusps also in $\eta\to3\pi^0$ might become feasible.
It has already been explored theoretically (along with $K_L\to3\pi^0$) in the non-relativistic
framework~\cite{bisseggercusp,bisseggerradcorr}
(for a qualitative study using a relativistic field theory, see Ref.~\cite{belina},
for an assessment in the framework of unitarized ChPT, see Ref.~\cite{nissler}).
As the present calculation is performed in chiral perturbation theory, 
and hence includes the rescattering effect leading to the cusp only at 
leading order in the quark mass expansion,
it is not suited to serve for a precision extraction of pion--pion scattering lengths,
but will rather illustrate the phenomenon.  
It is in a sense dual to the non-relativistic calculation as it aims at 
predicting electromagnetic effects in those parts of the amplitude
that are merely parameterized in the latter.

In Fig.~\ref{fig:dalitz} we show the cusp lines for the energy threshold of the production 
of two charged pions in all three kinematical channels $s$, $t$, and $u$ as a vertical, 
horizontal, and diagonal dashed line. 
Furthermore, for illustration purposes we plot the amplitudes along the lines $t=u$
in the following section; these lines are also drawn in Fig.~\ref{fig:dalitz}.

\subsection{Decay amplitudes}\label{sec:Namps}
\begin{sloppypar}
In this paper we focus on the electromagnetic contributions to $\eta\to3\pi$ decays 
and do not aim at a particularly reliable representation of the purely strong amplitude.
It is well known that one has to go beyond one-loop order to obtain
a valid representation of the latter~\cite{kambor,anisovich,Beisert,borasoy,bijnens},
while the $\order(\delta)$ part in our calculation corresponds precisely to 
the one-loop representation of Ref.~\cite{gleta}.  
This serves as a useful reference point to quantify the electromagnetic
corrections.  In the following, we only consider uncertainties in precisely these
electromagnetic contributions and disregard higher-order hadronic corrections.
Since the electromagnetic low-energy constants $K_i^r$ are not very well known,
we regard their input (and not unknown corrections of higher order in the chiral expansion)
as the dominant source of uncertainty; see Appendix~\ref{app:numbers} for a description
of how we vary the $K_i^r$.  
For some numbers we are forced to deviate from our standard procedure to estimate
the errors, those are marked by an asterisk in the following; the details are
also discussed in Appendix~\ref{app:numbers}.
All error bands in this section refer to this variation.
\end{sloppypar}

In the following, we compare the results for the one-loop amplitudes of $\order(\delta)$
(henceforth denoted by GL)~\cite{gleta}, those with effects of $\order(e^2)$ added (BKW)~\cite{bkw}, 
and the complete results of the present investigation, up-to-and-including 
effects of $\order(\delta e^2)$ (DKM).  
The GL and BKW amplitudes are also evaluated with our prescription
of the isospin limit for the pion mass 
(as well as with our choice of numerical input, Appendix~\ref{app:numbers})
in order to facilitate comparison with the higher-order corrections.

\begin{figure}
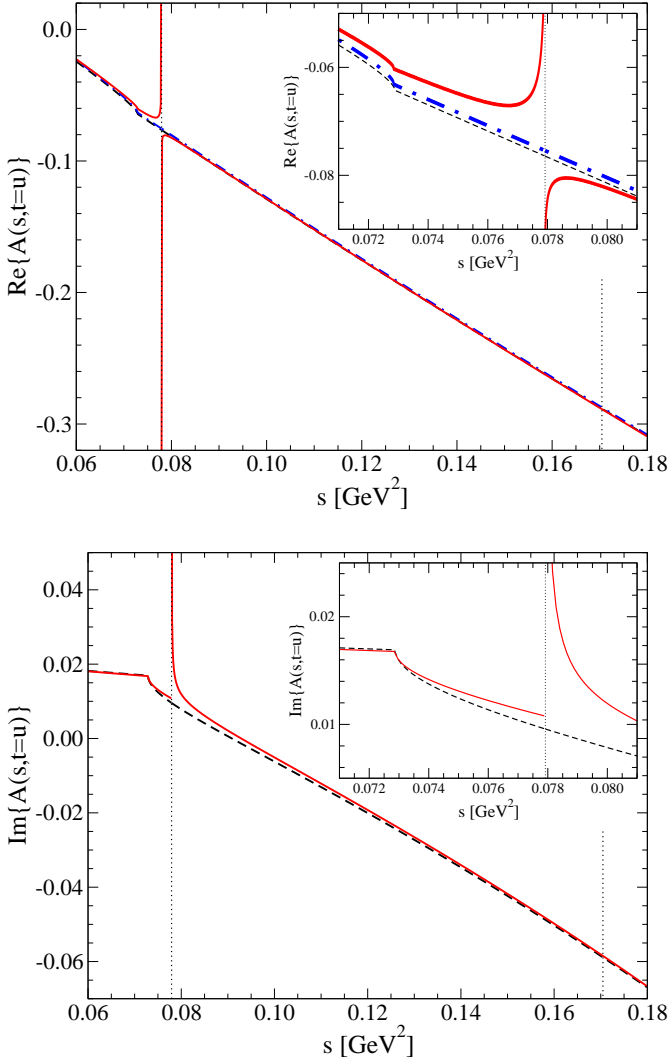

 \centering
 \includegraphics[width=\linewidth]{campre.eps}\\[5mm]
 \includegraphics[width=\linewidth]{campim.eps}
 \caption{Real and imaginary parts of the charged amplitudes GL (dashed/black), 
BKW (dot-dashed/blue), and DKM (full/red) for $t=u$.  
The inserts show the region close to the two-pion thresholds.
The line widths in the real part indicate the error bands due to 
a variation of electromagnetic low-energy constants.
The vertical lines show the limits of the physical region.} \label{fig:campzo}
\end{figure}
In Fig.~\ref{fig:campzo} we separately display the
real and the imaginary parts of the charged amplitudes GL, BKW, DKM
along the line $t=u$.
The infrared divergences in the amplitude are cured by hand according to~\eqref{eqn:IRcure}; 
the kinematical singularities at $s=4\mpi$ are retained here for illustration.
By our choice of the isospin limit for the pion mass,
the threshold cusp in the GL amplitude is artificially removed from the physical threshold energy 
to $s=4\mpn$.
According to~\eqref{eqn:cloamp}
the leading-order  charged decay amplitude is linear in $s$, 
which still dominates the energy dependence at NLO.
The BKW contributions are purely real for both decay channels, 
as no pion rescattering diagrams contribute at that particular order and
thus the imaginary parts of GL and BKW coincide with each other for both decays.
Furthermore, all imaginary parts are independent of any low-energy constants 
and are therefore plotted without an error range.

Since it is hard to identify cusps or the Coulomb pole in plots over the full kinematically allowed range, 
Fig.~\ref{fig:campzo} also contains inserts where we show 
the amplitudes close to the two-pion thresholds.
Here the expected features in the amplitudes are clearly visible: the $\pi^0\pi^0$ cusp 
at $4\mpn$ outside the physical region, as well as the Coulomb pole and phase divergence
at the $\pi^+\pi^-$ threshold.

\begin{figure}
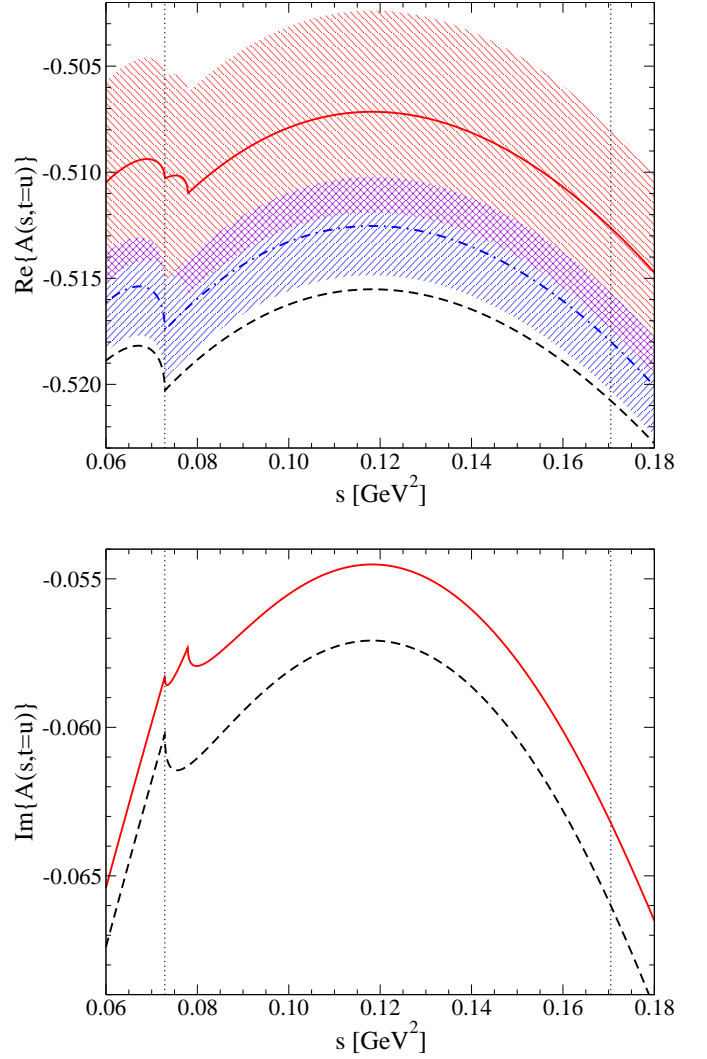

 \centering
 \includegraphics[width=\linewidth]{nampre.eps}\\[5mm]
 \includegraphics[width=\linewidth]{nampim.eps}
 \caption{Real and imaginary parts of the neutral amplitudes GL (dashed/black), 
BKW (dot-dashed/blue), and DKM (full/red) for $t=u$.  
The hatched regions denote the error bands due to 
a variation of electromagnetic low-energy constants.
The vertical lines show the limits of the physical region.} \label{fig:namp}
\end{figure}
The numerical results for the neutral decay amplitudes GL, BKW, and DKM
are shown in an analogous way in Fig.~\ref{fig:namp}. 
The vertical dotted lines indicate the physical region in $s$ for $t=u$
for the neutral decay, hence the lower bound is now at the $\pi^0\pi^0$ threshold. 
The leading decay amplitude for $\eta\to3\pi^0$ is constant, see~\eqref{eqn:nloamp},
and also at NLO the dependence on $s$ is weak.
Figure~\ref{fig:namp} shows that the size of the new contributions in DKM
is comparable to or even larger than those in BKW.
We find the expected cusp at the energy of the 
$\pi^+\pi^-$ threshold inside the physical region, although the overall variation is very small. 
As for the comparable decay $K_L \to 3\pi^0$,
the strength of the cusp might be rather small for a very precise determination and an extraction 
of $\pi\pi$ scattering lengths thereby.

\begin{sloppypar}
We wish to point out, however, that the cusp strength is underestimated here. 
As we mentioned before it is proportional to the combination of scattering lengths $a_0-a_2$, 
and the present ChPT calculation corresponds to the LO values 
$(a_0-a_2)^\tn{LO}=9\mpi/(32\pi F_\pi^2)=0.204$~\cite{weinberga0}\footnote{To be more precise,
the cusp strength in our calculation is determined by the leading-order scattering length
for $\pi^+\pi^- \to \pi^0\pi^0$ including isospin breaking, which is given by
$(a_0-a_2)^\tn{LO}\times\{1+(\mpi-\mpn)/(3\mpi)\}$, increasing the isospin-symmetric
value by about 2\%.}
while the matching of the Roy equations solution to the next-to-next-to-leading chiral representation
yields $(a_0-a_2)^\tn{NNLO}=0.265\pm0.004$~\cite{CGL,CGL2}, 
increasing the cusp strength considerably.
Furthermore, it has been noted before that due to the relative strength of the decays into 
charged and neutral final states, the cusp is much less pronounced in $\eta \to 3\pi^0$
than in $K^+ \to \pi^0\pi^0\pi^+$, much as in $K_L \to 3\pi^0$~\cite{bisseggercusp}. 
For these two reasons, we refrain from displaying the decay spectra $d\Gamma/ds$,
which would be the preferred observable for an extraction of $\pi\pi$ scattering lengths.
\end{sloppypar}

From Figs.~\ref{fig:campzo} and~\ref{fig:namp} we can conclude that the relative sizes of the 
electromagnetic corrections in the BKW and DKM amplitudes with respect to 
the strong result GL are of comparable size,
and the estimate of the electromagnetic effects, neglecting $m_u\neq m_d$ therein,
is not a very accurate representation.
However, the conclusion that the overall electromagnetic contributions remain very small
is still valid.

\subsection{Dalitz plot parameters}\label{sec:Dalitz}

The Dalitz plot distribution for the charged decay channel is described in terms of the symmetrized coordinates
\begin{align}
 x&=\sqrt{3}\frac{T_+-T_-}{Q_c}=\frac{\sqrt{3}(u-t)}{2M_\eta Q_c}\;,\nonumber
 \\
 y&=\frac{3T_0}{Q_c}-1=\frac{3\bigl[(M_\eta-M_{\pi^0})^2-s\bigr]}{2M_\eta Q_c}-1\;,
\end{align}
where $Q_c=T_0+T_++T_-=M_\eta-2M_\pi-M_{\pi^0}$ and the $T_i$ are the kinetic energies of the respective
pions in the $\eta$ rest frame.
For the neutral decay channel it is convenient to use one fully symmetrized coordinate
\begin{equation}
 z=\frac{2}{3}\sum_{i=1}^{3}\biggl(\frac{3T_i}{Q_n}-1\biggr)^2 
=x^2+y^2\;,
\end{equation}
with $Q_n = M_\eta-3M_{\pi^0}$, 
in order to reflect the symmetry in all Mandelstam variables.

\begin{sloppypar}
The squared absolute values of the two decay amplitudes are expanded around 
the center of the corresponding Dalitz plot according to the standard parameterizations
\begin{align}\label{eqn:dalitzamps}
 |\mathcal{A}_c(x,y)|^2 &= |\mathcal{N}_c|^2\bigl\{1+ay+by^2+dx^2 \nonumber\\
 & \qquad\qquad +fy^3+gx^2y+...\bigr\}\;,\nonumber
 \\
 |\mathcal{A}_n(x,y)|^2 &= |\mathcal{N}_n|^2\bigl\{1+2\alpha z+...\bigr\}\;,
\end{align}
in order to obtain the Dalitz slope parameters. 
For the charged channel odd terms in $x$ are forbidden due to charge conjugation symmetry. 
The cubic parameters have only been measured by the KLOE Collaboration~\cite{cKLOEnew}
with only $f$ found to be different from zero with statistical significance.
(For older experiments, see Refs.~\cite{Gormley,Layter,cCrystalBarrel}.)
\end{sloppypar}

Note that at second order in isospin breaking one has to take care of the precise definition 
of the center of the Dalitz plot. For the charged decay the point $s=t=u=s_0^c$ does not 
completely coincide with the point $x=y=0$ (where all kinetic energies are equal) 
due to the charged-to-neutral pion mass difference, see Ref.~\cite{bijgass}. 
Indeed,
$s=t=u=s_0^{c/n}$ corresponds to $x=y=0$ for the neutral decay, but $x=0$ and $y=0.0514$ 
for the charged decay.

We fit the distribution functions~\eqref{eqn:dalitzamps} to discretized grids 
(of roughly $200\times 200$ points)
of squared amplitudes over the whole physical region.
We use uniform weighting of the ``data'' points.
In order to quantify the fit quality for the amplitudes GL, BKW, DKM,
we normalize a fictitious $\chi^2$ to 1 for the GL amplitudes 
and only regard the relative changes, with the electromagnetic effects
switched on successively.
As explained above, in order to estimate the errors in the corrections to
the various Dalitz plot parameters, we vary the electromagnetic low-energy constants $K_i$
as described in Appendix~\ref{app:numbers}.  The resulting errors are always
larger than the fit errors on the extracted parameters, hence we can neglect the latter.  

\begin{table*}
 \centering
 \caption{Normalization and Dalitz plot parameters for $\eta\to\pi^+\pi^-\pi^0$ for the GL~\cite{gleta}, 
BKW~\cite{bkw}, and DKM (present work) amplitudes.  
All electromagnetic corrections are given as shifts 
(both absolute and relative, in percent) with respect to the strong result (GL).
For details, see main text. For a discussion of the errors, see Appendix~\ref{app:numbers}.}
 \medskip
 \newcommand{\tbk}{\hspace{-4mm}}
 \newcommand{\tpm}{\tbk$\pm$}
 \renewcommand{\arraystretch}{1.4}
 \begin{tabular}{crclrclrclc}\hline
      & \multicolumn{3}{c}{$|\mathcal{N}_c|^2$}&\multicolumn{3}{c}{$a$}&\multicolumn{3}{c}{$b$} &\\
\hline
  GL  & $ 0.0325$&    &              & $-1.279$&    &             & $0.396$&    &             & \\
  BKW & $-0.0004$&\tpm&\tbk$0.0003 $ & $-0.008$&\tpm&\tbk$0.001 $ & $+0.006$&\tpm&\tbk$0.001 $ & \\
      & $=(-1.1 $&\tpm&\tbk$0.9)\% $ & $=(+0.6$&\tpm&\tbk$0.1)\%$ & $=(+1.4$&\tpm&\tbk$0.2)\%$ & \\
  DKM & $-0.0008$&\tpm&\tbk$0.0002^* $ & $-0.009$&\tpm&\tbk$0.005 $ & $+0.006$&\tpm&\tbk$0.003 $ & \\
      & $=(-2.4 $&\tpm&\tbk$0.7^*)\%$& $=(+0.7$&\tpm&\tbk$0.4)\%$ & $=(+1.5$&\tpm&\tbk$0.7)\%$ & \\
\hline
      &\multicolumn{3}{c}{$d$} &\multicolumn{3}{c}{$f$}&\multicolumn{3}{c}{$g$}& $\chi^2/\ndf$ \\
\hline
  GL  & $ 0.0744$&    &              & $0.0126$&     &               & $-0.0586$&    &              & $\equiv1$\\
  BKW & $+0.0011$&\tpm&\tbk$0.0004 $ & $-0.0003$&\tpm&\tbk$0.0001  $ & $-0.0010$&\tpm&\tbk$0.0003 $ & $1.03$\\
      & $=(+1.5$ &\tpm&\tbk$0.5)\%$  & $=(-2.2$&\tpm&\tbk$0.4)\%$    & $=(+1.7$ &\tpm&\tbk$0.6)\%$  & \\
  DKM & $+0.0033$&\tpm&\tbk$0.0003^*$& $+0.0001$&\tpm&\tbk$0.0001  $ & $-0.0038$&\tpm&\tbk$0.0009^* $& $1.63$\\
      & $=(+4.4$ &\tpm&\tbk$0.4^*)\%$& $=(+0.5$&\tpm&\tbk$0.6)\%$    & $=(+6.4$ &\tpm&\tbk$1.5^*)\%$& \\
\hline
 \end{tabular}
 \renewcommand{\arraystretch}{1.0}
 \label{tab:cdalitzpar}
\end{table*}

The results for the normalization and the Dalitz plot parameters $a-g$ of the 
charged decay channel are shown in Table~\ref{tab:cdalitzpar}.  
We display the fit results for the GL, BKW, and DKM amplitudes.
Note again that the universal radiative corrections producing kinematical singularities
have been subtracted in the DKM amplitude according to the prescription given in Sect.~\ref{sec:universal}.

Table~\ref{tab:cdalitzpar} shows that, in general, electromagnetic corrections affect the Dalitz plot parameters
at the percent level.  The normalization tends to get reduced compared to the purely strong amplitude,
while the various slope parameters $a-g$ are slightly increased.  
The \emph{relative} shifts in $d$ and $g$ are more sizeable than in particular in $a$ and $b$ for the reason that 
the strong contribution to the $x$-dependent Dalitz plot parameters is suppressed to 
next-to-leading order in the chiral expansion
(as is obvious from the purely $s$-dependent tree-level amplitude~\eqref{eqn:cloamp}).
While the overall effects are  still very small,
we note that, throughout, the corrections of $\order(e^2)$ do not represent a valid estimate of the 
dominant electromagnetic corrections, as those of $\order(\delta e^2)$ are of the same order of magnitude --
sometimes with the same sign; sometimes both effects tend to cancel.  
We quote the corrections both as shifts of absolute size and as relative shifts in percent, compared
to the strong one-loop result~\cite{gleta}.
Regarding the fact that the strong corrections at $\order(p^6)$ are potentially substantial~\cite{bijnens},
the comparison of both 
may be taken as another indication of higher-order terms in the electromagnetic corrections.
The quality of the polynomial Dalitz plot fit is comparable throughout.

As explained before, the uncertainties quoted in Table~\ref{tab:cdalitzpar} are obtained by
varying the electromagnetic low-energy constants $K_i$.  
In most cases, the errors given are of the same order as the central shifts.

A few comments concerning the cubic Dalitz plot parameters $f$ and $g$ are in order.
In general, $f$ is highly correlated with $a$ in the fits.
In the electromagnetic corrections, we observe that the influence of the low-energy constants
on $f$ and $g$ is rather small, which is obvious from the fact that the counterterms only
lead to terms constant and linear in $s$.  We therefore do not consider the electromagnetic 
shifts thus obtained very reliable.
In particular, all errors due to the $K_i$ in the parameters $d$, $f$, $g$ 
are only indirectly induced by the variation in the normalization
and by the implicit inclusion of higher-order effects in the squared amplitude.
Note that, without subtracting the kinematical singularities due to universal corrections
(see Sect.~\ref{sec:universal}), in particular the fit results for $f$ become nonsensical.

\begin{table}
 \centering
 \caption{Normalization and Dalitz plot parameters for $\eta\to3\pi^0$ for the GL~\cite{gleta}, 
BKW~\cite{bkw}, and DKM (present work) amplitudes.  Results for the latter are also shown with 
the fit of the Dalitz plot region restricted as to exclude the cusps at the charged-pion thresholds
(DKM(cusp)).   All electromagnetic corrections are given as shifts 
(both absolute and relative, in percent) with respect to the strong result (GL).
For details, see main text.}
 \medskip
 \newcommand{\tbk}{\hspace{-4mm}}
 \newcommand{\tpm}{\tbk$\pm$}
 \renewcommand{\arraystretch}{1.4}
 \begin{tabular}{crclrclc}\hline
            &\multicolumn{3}{c}{$|\mathcal{N}_n|^2$}&\multicolumn{3}{c}{$10^2\times\alpha$}& \hspace{-3mm}$\chi^2/\ndf$\hspace{-2mm}\\
\hline
  GL        & $ 0.269$&    &             & $ 1.27$ &    &             & $\equiv1$\\
  BKW       & $-0.003$&\tpm&\tbk$0.002 $ & $+0.05$ &\tpm&\tbk$0.01$   & $0.99$   \\
            & $=(-1.1$&\tpm&\tbk$0.9)\%$ & $=(+3.7$&\tpm&\tbk$0.5)\%$ &          \\
  DKM       & $-0.009$&\tpm&\tbk$0.005 $ & $-0.002$&\tpm&\tbk$0.01$   & $6.20$   \\
            & $=(-3.3$&\tpm&\tbk$1.8)\%$ & $=(-0.2$&\tpm&\tbk$1.0)\%$ &          \\
\hspace{-2mm} DKM(cusp) \hspace{-3mm} 
            & $-0.009$&\tpm&\tbk$0.005 $ & $+0.06$ &\tpm&\tbk$0.01$   & $0.35$   \\
            & $=(-3.3$&\tpm&\tbk$1.8)\%$ & $=(+5.0$&\tpm&\tbk$1.1)\%$ &          \\
\hline
 \end{tabular}
 \renewcommand{\arraystretch}{1.0}
 \label{tab:ndalitzpar}
\end{table}
We now turn to the neutral channel $\eta\to3\pi^0$.  The results for the normalization
and the Dalitz plot parameter $\alpha$ are collected in Table~\ref{tab:ndalitzpar}.  
Here, in addition to the results using the GL, BKW, and DKM amplitudes, we also discuss
a variant of our result that takes into account that the DKM amplitude displays features
incompatible with a simple polynomial fit: the cusps at the charged-pion thresholds.
By DKM(cusp) we denote a fit to the part of the Dalitz plot with $z\leq z_0$, $z_0=0.598$ chosen  
such that the border region from the cusp outward is excluded.\footnote{Note that $z_0$ corresponds
to the \emph{minimum} value of $z$ where the cusps occur, as these show up for constant
$s$/$t$/$u$, not for constant $z$; see also the discussion in Ref.~\cite{nissler}.}
As for the charged decay channel, the normalization is reduced by electromagnetic corrections by a few percent.
Corrections of $\order(\delta e^2)$ are even bigger than those of $\order(e^2)$.  
We note that the cusp effect leads to the single biggest modification of any Dalitz plot parameter: 
trying to fit the cusp with the polynomial distribution function reduces $\alpha$ by 4\% (compare DKM to BKW
in Table~\ref{tab:ndalitzpar}), while excluding the cusp region increases it again by more than 5\%.
The latter shift is in qualitative agreement with the finding in Ref.~\cite{Deandrea}, where $\alpha$ is
determined just from the curvature at the center of the Dalitz plot.
The significance of this non-analytic structure is also reflected in the fit quality as quantified
by the $\chi^2/\ndf$ values quoted in Table~\ref{tab:ndalitzpar}: with the cusp included, the fit
becomes worse by a factor of 6 (DKM), while excluding it makes it even better than the fit of the GL 
distribution (DKM(cusp)).
Taking into account that the cusp strength is underestimated by about 30\%, see the discussion 
in Sect.~\ref{sec:Namps}, these numbers should be scaled accordingly.  
On the other hand, the cusp effect is by far too weak to contribute significantly to 
an explanation of the long-standing sign discrepancy for $\alpha$ between 
ChPT calculations~\cite{gleta,bijnens} and experimental 
determinations~\cite{GAMS2000,nCrystalBarrel,CrystalBall,SND,nKLOEnew,CELSIUSWASA,WASAatCOSY,unverzagtPhD,unverzagt}.\footnote{Note
that at $\order(p^6)$~\cite{bijnens}, $\alpha$ can be made consistent with experimental results
by particular choices of the low-energy constants.}
Uncertainties due to the $K_i$ are on the 1\% level throughout.

\begin{figure}
 \centering
\includegraphics[width=\linewidth]{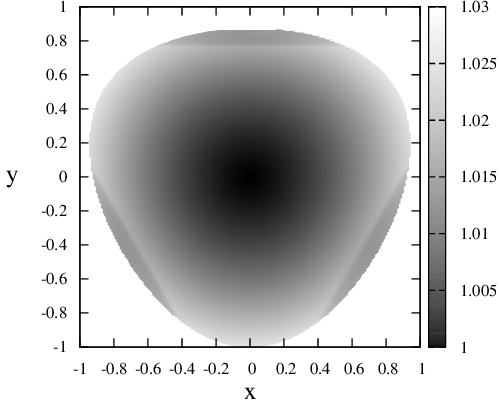}
 \caption{$\eta\to3\pi^0$ Dalitz plot distribution corresponding to the full one-loop amplitude (DKM),
including the cusp structures at the $\pi^+\pi^-$ thresholds.} \label{fig:3Dcusp}
\end{figure}
The $\eta\to3\pi^0$ Dalitz plot is displayed graphically for our central result, 
the full one-loop amplitude including electromagnetic corrections up to $\order(\delta e^2)$, 
in Fig.~\ref{fig:3Dcusp}.  
The cusp structures at the $\pi^+\pi^-$ thresholds are clearly visible.
Figure~\ref{fig:3Dcusp} also demonstrates that decay spectra with respect to $s$, $t$, $u$
are most sensitive to the cusp effect, not with respect to the radial coordinate $z$.

\subsection{Decay widths, branching ratio, quark mass ratios}

\begin{table}
 \centering
 \caption{Decay widths $\Gamma_{c/n}$ for both charged and neutral decay channel.  
The corrections for the BKW and DKM amplitudes are expressed relative to the strong GL result.
DKM(uc) denotes the variant of the amplitude without the universal photon corrections subtracted.}
 \medskip
 \newcommand{\tbk}{\hspace{-4mm}}
 \newcommand{\tpm}{\tbk$\pm$}
 \renewcommand{\arraystretch}{1.4}
 \begin{tabular}{crclrcl}\hline
   & \multicolumn{3}{c}{$\eta\to\pi^+\pi^-\pi^0$} & \multicolumn{3}{c}{$\eta\to3\pi^0$} \\\hline
  $\Gamma^\tn{GL}$ & \multicolumn{3}{l}{\hspace{4mm}$154.5\ev$} & \multicolumn{3}{l}{\hspace{4mm}$222.8\ev$} \\
  $\Delta\Gamma^\tn{BKW}$    & $ (-1.5$&\tpm&\tbk$1.3)  \ev$& $ (-2.5$&\tpm&\tbk$2.0)\ev$ \\
                             & $=(-1.0$&\tpm&\tbk$0.9)\%$   & $=(-1.1$&\tpm&\tbk$0.9)\%$ \\
  $\Delta\Gamma^\tn{DKM}$    & $ (-2.9$&\tpm&\tbk$0.7^*)\ev$& $ (-7.3$&\tpm&\tbk$4.0)\ev$ \\
                             & $=(-1.9$&\tpm&\tbk$0.5^*)\% $& $=(-3.3$&\tpm&\tbk$1.8)\%$ \\
  $\Delta\Gamma^\tn{DKM(uc)}$ & $ (-1.5$&\tpm&\tbk$0.7^*)\ev$& & \\
                             & $=(-1.0$&\tpm&\tbk$0.5^*)\% $& & \\\hline
 \end{tabular}
 \renewcommand{\arraystretch}{1.0}
 \label{tab:decaywidths}
\end{table}

The charged and neutral total decay widths can be calculated via
\begin{equation}
 \Gamma_{c/n}=\frac{S_{c/n}}{256\pi^3M_\eta^3}\int_{s_{c/n}^\tn{min}}^{s_{c/n}^\tn{max}}ds\int_{t_{c/n}^\tn{min}(s)}^{t_{c/n}^\tn{max}(s)}dt\,
\bigl|\amp_{c/n}(s,t,u)\bigr|^2\;,
\end{equation}
where the symmetry factor $S_n=1/3!$ for the neutral decay accounts for the indistinguishable 
three neutral pions, while $S_c=1$ for the charged decay. 
The $s$-integration boundaries are given by $s_c^\tn{min}=4\mpi$, $s_n^\tn{min}=4\mpn$, 
and $s_{c/n}^\tn{max}=(M_\eta-M_{\pi^0})^2$, whereas the $s$-dependent limits for the $t$-integration 
can be found in Ref.~\cite{pdg}.

The widths thus calculated are shown in Table~\ref{tab:decaywidths}.
The strong widths calculated at one loop underestimate the experimental values significantly,
so again,
the electromagnetic corrections are given as relative corrections to the strong result (GL).
For the charged decay the width is reduced moderately by about 2\%. 
We also show a variant of the DKM amplitude (denoted by DKM(uc)) where the universal photon
corrections have not been subtracted, i.e.\ the Coulomb pole at $s=4\mpi$ and the 
kinematical bremsstrahlung singularity at $s=(M_\eta-M_{\pi^0})^2$ are included.
In this case, the reduction of the width is only about 1\%.
In the neutral channel, the shifts in the width are completely dominated by the corrections
to the normalization of the amplitude, see Table~\ref{tab:ndalitzpar}.  
The total corrections up to $\order(\delta e^2)$ are about a factor of 3 larger than those estimated
by BKW.  The total reduction of the width by electromagnetic effects is in good agreement with
the one found in Ref.~\cite{Deandrea}.

\begin{table}
 \centering
 \caption{Branching ratios $r=\Gamma_n/\Gamma_c$.  The corrections for the 
BKW and DKM amplitudes are expressed relative to the strong GL result.
DKM(uc) denotes the variant of the amplitude without the universal photon corrections subtracted.}
 \medskip
 \renewcommand{\arraystretch}{1.4}
 \begin{tabular}{>{$}c<{$}>{$}c<{$}}\hline
   r^\tn{GL}            &  1.442 \\
  \Delta r^\tn{BKW}     & (-0.1\pm1.2)\%  \\
  \Delta r^\tn{DKM}     & (-1.4\pm1.8)\%  \\
  \Delta r^\tn{DKM(uc)} & (-2.3\pm1.8)\%  \\ 
\hline
 \end{tabular}
 \renewcommand{\arraystretch}{1.0}
 \label{tab:branchingratios}
\end{table}
From these widths one can determine the branching ratio
\begin{equation}
 r = \frac{\Gamma_{\eta\to3\pi^0}}{\Gamma_{\eta\to\pi^+\pi^-\pi^0}}\;
\end{equation}
at the different orders in isospin breaking.
The results are given in Table~\ref{tab:branchingratios}.
For the BKW corrections, as both widths are shifted by almost the same amount, the 
ratio is nearly constant compared to the strong result (GL).
Including all radiative corrections (DKM), $r$ is reduced by 1.4\%, 
with an uncertainty of about the same size. 
Without subtraction of the universal soft-photon corrections, the effect on $r$ 
is somewhat larger, about 2.3\%.

\begin{table}
 \centering
 \caption{Corrections to extractions of the quark mass ratio $Q$
from charged and neutral decay channel. For details, see main text.}
 \medskip
 \renewcommand{\arraystretch}{1.4}
 \begin{tabular}{>{$}c<{$}>{$}c<{$}>{$}c<{$}}\hline
                         & \eta\to\pi^+\pi^-\pi^0      & \eta\to3\pi^0   \\ \hline
   \Delta Q^\tn{BKW}     & (+0.24\pm0.22\phantom{^*})\%& (+0.28\pm0.22)\%\\
   \Delta Q^\tn{DKM}     & (+0.48\pm0.12^*)\%          & (+0.84\pm0.46)\%\\
   \Delta Q^\tn{DKM(uc)} & (+0.24\pm0.12^*)\%          &                 \\ \hline
 \end{tabular}
 \renewcommand{\arraystretch}{1.0}
 \label{tab:Q}
\end{table}
Finally, we can read off the corrections for the quark mass ratio $Q$
from the relation
\beq
\Gamma_{c/n} ~\propto~ \frac{1}{Q^4} ~. \label{eqn:GammaQ4}
\eeq
Strictly speaking, the relation~\eqref{eqn:GammaQ4} does not hold for the electromagnetic
terms of $\order(e^2(m_u=m_d))$ as present in the BKW corrections.  However
given the smallness of the corrections and the size of the uncertainties, the additional
error in factorizing $Q^{-2}$ in the complete amplitude can be safely neglected.
The resulting shifts in $Q$ are collected in Table~\ref{tab:Q}.  
They are to be interpreted in the following way: extracted naively according 
to~\eqref{eqn:GammaQ4}, $Q$ would be shifted compared to the purely strong value
according to Table~\ref{tab:Q}, i.e.\ generally increased.
In order to purify a real measurement (that includes electromagnetic effects) such
that the purely strong value for $Q$ is obtained, the opposite shift has to be applied.
The electromagnetic corrections in Table~\ref{tab:Q} are below half a percent
for the charged channel, while in the neutral channel, a value of $Q$ would be reduced
by close to 1\%.

\setcounter{figure}{0}
\setcounter{table}{0}
\section{Conclusions}\label{sec:conclusions}

\begin{sloppypar}
We have re-evaluated electromagnetic contributions to $\eta\to3\pi$ decays,
thus calculating the corrections to Sutherland's theorem~\cite{sutherland,bell},
extending and updating an earlier analysis~\cite{bkw} that neglected terms of $\order(e^2(m_u-m_d))$.
Terms of this order contain the essential non-analytic structures affecting the
Dalitz plot distribution, in particular real- and virtual-photon corrections
in $\eta\to\pi^+\pi^-\pi^0$, and the cusps due to virtual $\pi^+\pi^-$ intermediate states
in $\eta\to3\pi^0$.  
We have shown (explicitly at the soft-pion point) 
that these additional terms are not suppressed by an additional small isospin-breaking
parameter, but only by a factor of $(m_d-m_u)/(m_d+m_u) \approx 1/3$.
Terms of $\order(e^2(m_u-m_d))$ violate the leading isospin relation between charged and neutral decay 
amplitudes.

We have calculated the corrections to the various Dalitz plot parameters, 
as well as for the widths and branching ratios, and the resulting correction
for an extraction of the quark mass ratio $Q$.  
Although the effects of $\order(e^2(m_u-m_d))$ 
are of the same size as those analyzed in Ref.~\cite{bkw} and are not any
further suppressed, the total electromagnetic corrections remain very small, 
at the percent level, throughout.  The most significant change occurs in the
$\eta\to3\pi^0$ Dalitz plot parameter $\alpha$, which is slightly more strongly
affected by the presence of the cusps.  The latter are, however, not sufficient to
explain the sign discrepancy between chiral predictions and experimental results.
\end{sloppypar}

The present study does not replace a dedicated analysis of the cusp effect in $\eta\to3\pi^0$
in non-relativistic field theory~\cite{bisseggercusp,bisseggerradcorr,gullstrom}, 
but rather complements it in also predicting the polynomial electromagnetic terms.
As such, our results ought to serve as correction factors in future precision
experiments on $\eta\to3\pi$ decays.

\begin{acknowledgement}
\begin{sloppypar}
\textit{Acknowledgements.}
We thank Hans Bijnens for useful discussions and e-mail communications,
and V\'eronique Bernard for useful information on the value of $F_K/F_\pi$.
Partial financial support by the Helmholtz Association through funds provided
to the virtual institute ``Spin and strong QCD'' (VH-VI-231),
by the EU Integrated Infrastructure Initiative Hadron Physics Project 
(contract number RII3--CT--2004--506078),
and by DFG (SFB/TR 16, ``Subnuclear Structure of Matter'') is gratefully
acknowledged. 
\end{sloppypar}
\end{acknowledgement}

\begin{appendix}

\setcounter{figure}{0}
\setcounter{table}{0}
\section{\boldmath{$\eta\to3\pi$} amplitudes}\label{app:amps}

In this appendix, we collect the explicit formulae for the full $\eta\to3\pi$ decay amplitudes.
In writing these down, we make use of the 
relation
\begin{equation}
\epsilon = \frac{B_0\delta}{\sqrt{3}(\me-\mpn)} = \frac{\sqrt{3}(3\me+\mpn)}{16\mpn Q^2} \label{eqn:epsdelQ2} ~,
\end{equation}
valid at leading chiral order and to leading order in isospin breaking,
and use the redundant quantities $\epsilon$, $\delta$, $Q^{-2}$ to express the various
next-to-leading order contributions in a more compact form, where possible.  
For numerical evaluations, only $Q$ is employed, hence $\epsilon$ and $\delta$ are always to be 
taken as abbreviations according to \eqref{eqn:epsdelQ2}.
In a similar spirit, we do not perform the expansion of kaon loop functions according 
to~\eqref{eqn:kaontadexp} and \eqref{eqn:kaonloopexp} explicitly, as these also
tend to make the results more cumbersome.

\subsection{Renormalization factors, mixing}\label{app:renorm}

First we write down the $Z$ factors for the wave-function renormalization that follow from the full propagators at NLO 
shown in Fig.~\ref{fig:ffullprop} and the renormalization corrections defined in~\eqref{eqn:rencorr}. 
As the leading-order amplitudes for both $\eta\to\pi^+\pi^-\pi^0$ and $\eta\to3\pi^0$ are of $\order(\delta)$, 
all formulae are only needed up to $\order(e^2)$ in isospin breaking. The $Z$ factors read
\begin{align}
 Z_\eta&=1+\frac{8}{F_\pi^2}\biggl\{\frac{\Delta_K\!+\!\Delta_{K^0}}{16}
-\frac{3}{2}(\me+\mpn)L_4-\me L_5\biggr\}\nonumber\\
 &-\frac{4}{3}e^2\biggl\{2(K_1+K_2)-K_3+\frac{K_4}{2}+K_5+K_6\biggr\} \!+\! \order(\delta) \,,
 \\
 Z_{\pi^0}&=1+\frac{8}{F_\pi^2}\biggl\{\frac{\Delta_K+\Delta_{K^0}+4\Delta_\pi}{48}
-\frac{3}{2}(\me+\mpn)L_4 \nonumber\\
&\quad - \mpn L_5\biggr\} 
 -e^2\biggl\{\frac{8}{3}(K_1+K_2)-2(2K_3-K_4)\nonumber\\
&\quad +\frac{20}{9}(K_5+K_6)\biggr\}+\order(\delta)\;,
 \\
 Z_\pi&=1+\frac{8}{F_\pi^2}\biggl\{ \frac{\Delta_K+\Delta_{K^0}+2(\Delta_\pi+\Delta_{\pi^0})}{48} - \mpn L_5 
\nonumber\\
&\quad-\frac{3}{2}(\me+\mpn)L_4 \biggr\}
-e^2\biggl\{\frac{1}{8\pi^2}\biggl(1+\ln\frac{m_\gamma^2}{\mpi}\biggr)\nonumber\\
& \quad+2\frac{\Delta_\pi}{\mpi} + \frac{8}{3}(K_1+K_2) +\frac{20}{9}(K_5+K_6) 
\biggr\}+\order(\delta)\;.
\end{align}
In order to account for $\pi^0\eta$ mixing due to the diagrams depicted in Fig.~\ref{fig:fnlomix} 
at next-to-leading order we need the quantities $\epsilon_4$ and $Z_{\pi^0\eta}$ defined in~\eqref{eqn:mixdef},
\begin{align}
 \epsilon_4&=
-\frac{2B_0\delta}{\sqrt{3}F_\pi^2}\biggl\{12(3L_7+L_8)
+\frac{\mpn(\Delta_K\!+\!\Delta_{K^0}\!-\!2\Delta_\pi)}{3(\me-\mpn)^2} \nonumber\\
 &\quad-\frac{\Delta_\eta-\Delta_{\pi^0}}{4(\me-\mpn)}
\biggr\}
-\frac{(3\me+\mpn)(\Delta_K-\Delta_{K^0})}{4\sqrt{3}F_\pi^2(\me-\mpn)}
\nonumber\\
 &-\frac{2e^2}{3\sqrt{3}(\me-\mpn)}\biggl\{\biggl[\frac{3}{2}(2K_3-K_4)-K_5-K_6\biggr] \nonumber\\
&\qquad \times
(\me+\mpn)
\Bigl(1-\frac{2\epsilon}{\sqrt{3}}\Bigr)\nonumber\\
 &\quad+2(K_9+K_{10})\biggl(\mpn-\frac{\epsilon}{\sqrt{3}}(3\me-\mpn)\biggr)\biggr\} 
\nonumber\\&
+\order(\delta^2)\;, 
\\
 Z_{\pi^0\eta}&=\frac{\Delta_K-\Delta_{K^0}}{2\sqrt{3}F_\pi^2}
+\frac{\epsilon}{3F_\pi^2}\bigl\{\Delta_K+\Delta_{K^0}-2\Delta_\pi\bigr\} \nonumber\\
 &+\frac{2e^2}{3\sqrt{3}}\Bigl(1-\frac{2\epsilon}{\sqrt{3}}\Bigr)\Bigl\{3(2K_3-K_4) -2(K_5+K_6)\Bigr\} 
\nonumber\\&
+\order(\delta^2)\;,
\end{align}
that multiply the lowest-order amplitudes
\begin{align}
 \amp_{\pi^0\to\pi^+\pi^-\pi^0}^\tn{LO}&=\frac{3s-\me-2\mpn}{3F_\pi^2} \;, \nonumber\\
 \amp_{\eta\to\eta\pi^+\pi^-}^\tn{LO}&= \amp_{\eta\to\eta\pi^0\pi^0}^\tn{LO} = \frac{1}{3} \amp_{\pi^0\to3\pi^0}^\tn{LO}
= \frac{\mpn}{3F_\pi^2} \;,
\end{align}
where quark mass insertions have been replaced by leading meson masses since both $Z_{\pi^0\eta}$ and $\epsilon_4$ are 
next-to-leading order effects.

\begin{sloppypar}
The various renormalization corrections defined in Sect.~\ref{sec:tree}, as well as $\Delta_\tn{GMO}$
and $\Delta_F$ that we use to express strong low-energy constants in terms of observable quantities,
are given by
\begin{align}
 \Delta_\tn{GMO}&=\frac{1}{2F_\pi^2}\bigg\{12(\me-\mpn)\Bigl[L_5-6(2L_7+L_8)\Bigr]\nonumber\\
 &\quad-\frac{1}{\me-\mpn}\Bigl[(3\me+\mpn)(\Delta_K+\Delta_{K^0})\nonumber\\
 &\qquad-2\mpn(2\Delta_\pi-\Delta_{\pi^0})-6\me\Delta_\eta\Bigr]\bigg\}\nonumber\\
 &-2e^2\bigg\{3\Bigl[\frac{\Delta_K-\Delta_\pi}{\me-\mpn}-\frac{1}{16\pi^2}\Bigr]+2K_3-K_4\nonumber\\
 &\quad+\frac{1}{3}(K_5+K_6)-6(K_{10}+K_{11})\bigg\}\nonumber\\
 &-\frac{2\Delta\mpi}{F_\pi^2}\Bigl\{6L_5+\frac{\Delta_K-\Delta_\pi}{\me-\mpn}\Bigr\} +\order(\delta)\;,
\\
 \Delta_F&=\frac{1}{F_\pi^2}\bigg\{\frac{1}{8}\Bigl[5\Delta_{\pi^0}-2\Delta_{K^0}-3\Delta_\eta\Bigr]\nonumber\\
 &\quad+3(\me-\mpn)L_5\bigg\}+\order(e^2,\delta)\;, 
\\
 \Delta_M&=\frac{1}{6F_\pi^2}\bigl\{(3\me+\mpn)(\Delta_K+\Delta_{K^0}) \nonumber\\
 &\quad -2\mpn(4\Delta_\pi-\Delta_{\pi^0})-2(2\me-\mpn)\Delta_\eta\bigr\}\nonumber\\
 &+\frac{4(\me-\mpn)}{F_\pi^2}\Bigl\{2(\me-\mpn)(3L_7+L_8)\nonumber\\
 &\quad+(\me+\mpn)\Bigl[3(2L_6-L_4)-2(L_5-2L_8)\Bigr]\Bigr\}\nonumber\\
 &-\frac{4e^2}{3}\Bigl\{2(\me-\mpn)(K_1+K_2-K_7-K_8) \nonumber\\
& \quad-\frac{1}{2}(\me-3\mpn)(2K_3-K_4)\nonumber\\
 &\quad+\frac{1}{3}(3\me-5\mpn)(K_5+K_6)\nonumber\\
 &\quad-\frac{2}{3}(\me-2\mpn)(K_9+K_{10})\Bigr\}+\order(\delta)\,, 
 \\
 \Delta_Z&=-\frac{\mpn}{F_\pi^2}(\Delta_\pi-\Delta_{\pi^0})
-\frac{\Delta\mpi}{F_\pi^2}\Bigl\{2\Delta_\pi+\Delta_K \nonumber\\
&\quad +4\Bigl[3(\me+\mpn)L_4+2\mpn L_5\Bigr]\Bigr\}\nonumber\\
 &-e^2\Bigl\{2\mpn\Bigl[2K_3-K_4-4(K_{10}+K_{11})-\frac{1}{8\pi^2}\Bigr]\nonumber\\
&\quad -6(\me+\mpn)K_8 + 3\Delta_\pi \Bigr\}
+\order(\delta)\;,
 \\
 \Delta_Q&=\frac{1}{F_\pi^2}\biggl\{\frac{\Delta_K\!+\!\Delta_{K^0}}{2} -\Delta_\pi 
+\frac{3\me-\mpn}{3\me+\mpn}\frac{\Delta_{\pi^0}\!-\!\Delta_\eta}{2}\nonumber\\
 &-\frac{24(\me-\mpn)}{3\me+\mpn}\Bigl[\me (L_5-2L_8) \nonumber\\
& \qquad - (\me-\mpn)(3L_7+L_8)\Bigr]\biggr\}\nonumber\\
 &-\frac{4e^2}{3\me+\mpn}\Bigl\{\me\Bigl[2K_3-K_4-\frac{2}{3}(K_5+K_6)\Bigr]\nonumber\\
 &\quad+\frac{1}{3}(3\me-\mpn)(K_9+K_{10})\Bigr\}+\order(\delta)\;,
 \\
 \Delta_{F_\pi}&=\frac{1}{F_\pi^2}\Bigl\{6(\me+\mpn)L_4+4\mpn L_5
\nonumber\\& \qquad 
-\frac{\Delta_{K^0}}{2} -\Delta_{\pi^0}\Bigr\}+\order(e^2,\delta)\;.
\end{align}
\end{sloppypar}

\subsection{$\boldsymbol{\eta\to\pi^+\pi^-\pi^0}$ decay amplitude}\label{app:charged}

All loop contributions in the following are given up to the order in isospin breaking
considered in this work, i.e.\ $\order(\delta e^2)$.  
Summing up the tadpole diagrams depicted in Fig.~\ref{fig:fmesonloops}a) for the charged decay channel yields
\begin{align}
 \amp_c^\tn{tad}&=-\frac{(\Delta_K-\Delta_{K^0})(s-\me-\mpn)}{12\sqrt{3}F_\pi^4} \nonumber\\
&+\frac{\epsilon}{18F_\pi^4}\biggl\{ 
(5s+\me-3\mpn)(\Delta_K+\Delta_{K^0}) \nonumber \\
& \quad +(8s+\me-15\mpn)\Delta_{\pi^0} + (\me+\mpn)\Delta_\eta \nonumber \\
& \quad+4(3s-\me-5\mpn+4\Delta\mpi)\Delta_\pi 
\biggr\} \;.
\end{align}

Next we give the contributions from the $s$, $t$, $u$ channel loop diagrams Fig.~\ref{fig:fmesonloops}b) 
in terms of the loop functions defined in Appendix~\ref{app:loops}.
The superscripts $s$, $t$, $u$ of the amplitudes specify the channel, the subscripts 
for the $t$ channel diagrams the virtual mesons 
in the intermediate state. The $s$ channel contributions are given by
\begin{align}
 \amp_c^s &=\frac{1}{48\sqrt{3}F_\pi^4}\Big\{3s(3s\!-\!3\me\!-\!\mpn)\bigl(J_{KK}(s)\!-\!J_{K^0K^0}(s)\bigr)\nonumber\\
 &\quad+2(5s-3\me-\mpn)(\Delta_{K}-\Delta_{K^0})\Big\} \nonumber\\
 &+\frac{\Delta\mpi}{4\sqrt{3}F_\pi^4}\Big\{(3s-3\me-\mpn)J_{KK}(s)+2\Delta_{K}\Big\} \nonumber\\
 &+\frac{\epsilon}{6F_\pi^4}\Big\{(\me-\mpn)\Big[\mpn J_{\eta\eta}(s) \nonumber\\
 &\qquad-3(s-\mpn)J_{\pi^0\pi^0}(s)-2\Delta_{\pi^0}\Big] \nonumber\\
 &\quad+\frac{1}{12}\Big[(3s-4\mpn)\Big\{3sJ_{K^0K^0}(s)-8\mpn J_{\eta\pi^0}(s) \nonumber\\
 &\qquad\quad+3(s+4\Delta\mpi)(J_{KK}(s)-4J_{\pi\pi}(s))\Big\} \nonumber\\
 &\qquad+2(5s-4\mpn)\Delta_{K^0}-8\mpn(\Delta_{\eta}+\Delta_{\pi^0}) \nonumber\\
 &\qquad+2(5s-4\mpn+12\Delta\mpi)(\Delta_K-4\Delta_\pi)\Big]\Big\} \;.
\end{align}
The individual $t$ channel contributions can be written as
\begin{align}
& \amp^t_{\pi^0\pi} = -\frac{\epsilon}{18F_\pi^4}\biggl\{\frac{1}{16\pi^2}
\biggl[(t-6\mpn)(t+2s-\me-3\mpn)\nonumber\\
 &\quad-\biggl(5t\!+\!6s\!-\!2\me\!-\!2\mpn\biggl(11\!+\!3\frac{\me\!-\!\mpn}{t}\biggr)\biggr)\Delta\mpi\biggr]\nonumber\\
 &+3\biggl[2t^2+t(s-2\me-7\mpn)-\mpn(4s-5\me-7\mpn)\nonumber\\
 &\quad-\biggl(8t^2+2t(s-\me-8\mpn) 
-5\mpn(\me-\mpn)\biggr)\nonumber\\
& \qquad \times \frac{\Delta\mpi}{t}\biggr] J_{\pi^0\pi}(t)\nonumber\\
 &+3\biggl[\frac{4}{3}t+s-\me-\frac{8}{3}\mpn-\biggl(3-\frac{\me-\mpn}{t}\biggr)\Delta\mpi\biggr] \nonumber\\
& \qquad \times (\Delta_\pi+\Delta_{\pi^0})\nonumber\\
 &-3(\me-\mpn)(t-3\mpn)\frac{\Delta_\pi-\Delta_{\pi^0}}{t}\biggr\}\;,
\\
& \amp^t_{K^0K} =-\frac{\Delta\mpi}{8\sqrt{3}F_\pi^4}\biggl\{\bigl(3t-3\me-\mpn\bigr)J_{K^0K}(t)+\Delta_K\nonumber\\
 &\quad+\Delta_{K^0}-\biggl(2-\frac{3\me+\mpn}{t}\biggr)\bigl(\Delta_K-\Delta_{K^0}\bigr)\biggr\}\nonumber\\
 &-\frac{\epsilon}{24F_\pi^4}\biggl\{\frac{1}{48\pi^2}
\biggl[(t\!+\!2s\!-\!\me\!-\!3\mpn)(2t\!-\!9\me\!-\!3\mpn)\nonumber\\
 &\quad -\biggl(2(5t+6s-11\me-13\mpn) \nonumber\\
& \qquad -3\frac{(\me-\mpn)(3\me+\mpn)}{t}\biggr)\Delta\mpi\biggr]\nonumber\\
&-2\biggl[4t^2\!-\!t(s\!+\!7\me\!+\!\mpn) 
\!+\!(s\!+\!\me\!-\!\mpn)(3\me\!+\!\mpn)\nonumber\\
 &\quad-\biggl(7t-2s-\me-3\mpn \nonumber\\
& \qquad -\frac{(\me-\mpn)(3\me+\mpn)}{t}\biggr)
\Delta\mpi\biggr]J_{K^0K}(t)\nonumber\\
 &\quad-\biggl[4t-2s-5\me+\mpn \nonumber\\
& \qquad -2\biggl(2+\frac{\me-\mpn}{t}\biggr)\Delta\mpi\biggr]
(\Delta_K+\Delta_{K^0})\biggr\} \,, \hspace{-1mm}
\\
& \amp^t_{\eta\pi}=\frac{\epsilon  \mpn}{6F_\pi^4}\biggl\{ 
\frac{\Delta_\eta + \Delta_\pi}{3}
-\frac{\Delta\mpi}{t} \bigl(\Delta_\eta - \Delta_\pi \bigr) \nonumber\\
&+ \biggl[t-\me - \frac{\mpn}{3} 
-\biggl(2+\frac{\me-\mpn}{t}\biggr)\Delta\mpi\biggr]J_{\eta\pi}(t) \biggr\} \;.
\end{align}
The total $t$ channel contribution is therefore
\begin{equation}
 \amp_c^{tu}(s,t)\equiv \amp^t_{\pi^0\pi}+\amp^t_{K^0K}+\amp^t_{\eta\pi} \;.
\end{equation}
The $u$ channel graphs are given by crossing symmetry as $\amp_c^{tu}(s,u)$.

The triangle diagram shown in Fig.~\ref{fig:fphotonloops}b) gives rise to a contribution of the form
\begin{align}\label{eqn:triangleamp}
 \amp_c^{\pi\pi\gamma}&=\frac{e^2\epsilon}{3F_\pi^2}
\biggl\{(3s-4\mpn)\biggl[2(s-2\mpi)G(s)+J_{\pi\pi}(s) \nonumber\\
& \qquad +2\frac{\Delta_\pi}{\mpi}-\frac{1}{8\pi^2}\biggr] 
+(s-2\mpi)\biggl[3\frac{\Delta_\pi}{\mpi}-\frac{1}{4\pi^2}\biggr]\biggr\} \;,
\end{align}
and the vertex correction diagram depicted in Fig.~\ref{fig:fphotonloops}a) leads to
\begin{align}
 \amp_c^{\pi\gamma}&=-\frac{e^2\epsilon(s-4\mpi)}{2F_\pi^2}\biggl\{\frac{\Delta_\pi}{\mpi}
-\frac{1}{12\pi^2}\biggr\} \;.
\end{align}

The result of the various replacements in $\amp_c^\tn{tree}$ for the charged decay is given by the total tree amplitude
\begin{align}
 &\amp_c^\tn{tree}=\frac{\Delta_K-\Delta_{K^0}}{6\sqrt{3}F_\pi^4}
\biggl[\frac{(3s-4\mpn)2\me}{\me-\mpn}-(2\me-\mpn)\biggr]\nonumber\\
 &\quad+\frac{e^2(3s-4\mpn)2\mpn}{9\sqrt{3}F_\pi^2(\me-\mpn)}
\Bigl(1-\frac{2\epsilon}{\sqrt{3}}\Bigr)\nonumber\\
 &\quad\times\Bigl[3(2K_3-K_4)-2(K_5+K_6)+2(K_9+K_{10})\Bigr]\nonumber\\
 &-\frac{1}{Q^22\sqrt{3}F_\pi^2\mpn}\Biggl\{(3s-4\mpn)\Biggl[\frac{3\me+\mpn}{8}\nonumber\\
 &\qquad+\frac{\me}{2}\Delta_\tn{GMO}+\frac{(3\me+\mpn)\mpn+6\me\Delta\mpi}{3(\me-\mpn)}\Delta_{F}\nonumber\\
 &\qquad+\frac{1}{24F_\pi^2}\biggl((21\me+\mpn)(\Delta_\pi-\Delta_{\pi^0})
+ \frac{3\me+\mpn}{\me-\mpn}\nonumber\\
 &\qquad\quad\times\Bigl\{(11\me-3\mpn)\frac{\Delta_K}{2}+(5\me+7\mpn)\frac{\Delta_{K^0}}{2}\nonumber\\
 &\qquad\quad-3(3\me-\mpn)\Delta_\eta - (\me+\mpn)(\Delta_\pi+\Delta_{\pi^0})\Bigr\}\nonumber\\
 &\qquad\quad+\frac{6\Delta\mpi\me}{\me-\mpn}\biggl\{4\Bigl(1+\frac{3e^2F_\pi^2}{\Delta\mpi}\Bigr)
(\Delta_K-\Delta_\pi)\nonumber\\
 &\qquad\qquad+2\Delta_{K^0}-5\Delta_{\pi^0}+3\Delta_\eta\biggr\}\biggr)\Biggr]\nonumber\\
 &\quad-(3\me+\mpn)\Biggl[\frac{\mpn}{3}\Delta_\tn{GMO}+\frac{(\me+\mpn)\Delta\mpi}{\me-\mpn}\Delta_F\nonumber\\
 &\quad-\frac{L_3}{2F_\pi^2}\Bigl\{(s\!-\!\me)(s\!-\!2\Delta\mpi)
\!+\!2tu\!-\!(3s\!+\!2\Delta\mpi)\mpn\Bigr\}\nonumber\\
 &\quad+\frac{1}{12F_\pi^2}\biggl(\!
\frac{11}{2}\mpn(\Delta_K\!+\!\Delta_{K^0})
\!+\! (\me\!-\!4\mpn)(\Delta_\pi\!-\!\Delta_{\pi^0})  \nonumber\\
 &\qquad - \frac{1}{\me-\mpn}\biggl[4\mpn(3\me-\mpn)\Delta_\eta \nonumber\\
 &\qquad\quad - \frac{M_\eta^4+15M_{\pi^0}^4}{2} \Bigl( \Delta_K+\Delta_{K^0}-\Delta_\pi
                +\frac{\Delta_{\pi^0}}{2}  -\frac{\Delta_\eta}{2} \Bigr) \nonumber\\
 &\qquad\quad-3e^2F_\pi^2\bigl\{8\mpn\Delta_K-(3\me+5\mpn)\Delta_\pi\bigr\}\nonumber\\
 &\qquad\quad-\Delta\mpi\Bigr\{\frac{3}{2}(\me+\mpn)(2\Delta_{K^0}-5\Delta_{\pi^0}+3\Delta_\eta)\nonumber\\
 &\qquad\qquad-(3\me\!-\!11\mpn)\Delta_K
-2(3\me\!+\!\mpn)\Delta_\pi\Bigr\}\biggr]\biggr)\Biggr]\nonumber\\
 &\quad-e^2(3s-4\mpn)\Biggl[\frac{15\me+\mpn}{64\pi^2}+(3\me+\mpn)\nonumber\\
 &\qquad\times\biggl\{\frac{1}{64\pi^2}\ln\frac{m_\gamma^2}{\mpi}+\frac{\Delta_\pi}{4\mpi}+\frac{1}{3}(K_1+K_2+K_5)\biggr\}\nonumber\\
 &\qquad-\!\Bigl(\me\!-\!\frac{\mpn}{6}\Bigr)(2K_3\!-\!K_4)
\!-\!\me(K_9\!-\!5K_{10}\!-\!6K_{11})\Biggr]\nonumber\\
 &\quad+e^2(3\me+\mpn)\Biggl[\frac{\mpn}{16\pi^2}+\frac{3}{4}(\me-\mpn)(2K_3-K_4)\nonumber\\
 &\qquad-\frac{\mpn}{3}K_5-\me K_6+3(\me\!+\!\mpn)(K_{10}+K_{11})\Biggr]\Biggr\} \,.
\end{align}

\subsection{$\boldsymbol{\eta\to3\pi^0}$ decay amplitude}\label{app:neutral}

For the neutral decay channel, the tadpole diagrams depicted in Fig.~\ref{fig:fmesonloops}a) sum up to
\begin{align}
 \amp_n^\tn{tad}&=\frac{\me(\Delta_K-\Delta_{K^0})}{6\sqrt{3}F_\pi^4} 
 +\frac{\epsilon}{18F_\pi^4}\biggl\{ 3(\me+\mpn)\Delta_\eta \nonumber\\
&+2(\me-15\mpn)\Delta_\pi+3(3\me-5\mpn)\Delta_{\pi^0}\nonumber \\
&+ 2(4\me+3\mpn)(\Delta_K+\Delta_{K^0}) 
\biggr\} \;.
\end{align}

The rescattering diagrams displayed in Fig.~\ref{fig:fmesonloops}b) give rise to the following $s$ channel contributions:
\begin{align}
&\amp_n^{stu}(s)=\frac{1}{48\sqrt{3}F_\pi^4}\Bigl\{2(5s-3\me-\mpn)\bigl(\Delta_K-\Delta_{K^0}\bigr)\nonumber\\
 &\qquad\quad+3s(3s-3\me-\mpn)\bigl(J_{KK}(s)-J_{K^0K^0}(s)\bigr)\Bigr\}\nonumber\\
 &+\frac{\epsilon}{6F_\pi^4}\Bigl\{(\me\!-\!\mpn)\mpn\Big[J_{\eta\eta}(s)
-3J_{\pi^0\pi^0}(s)-2J_{\eta\pi^0}(s)\Big]\nonumber\\
 &\quad+\frac{1}{4}\Bigl[12s^2-2s(9\me+5\mpn)+(3\me+\mpn)^2\Bigr]\nonumber\\
 &\qquad\qquad \times\bigl(J_{KK}(s)+J_{K^0K^0}(s)\bigr)\nonumber\\
 &\quad+\frac{1}{3}(10s-9\me-5\mpn)\bigl(\Delta_K+\Delta_{K^0}\bigr)\nonumber\\
 &\quad-2(3s-4\mpn)(s-\mpn)J_{\pi\pi}(s)
-\frac{4}{3}(5s-7\mpn)\Delta_{\pi} \!\Bigr\} \,.
\end{align}
The crossed channels are given by $\amp_n^{stu}(t) + \amp_n^{stu}(u)$.

Replacing the strong LECs in $\amp_n^\tn{tree}$ for the neutral decay as explained in the main text 
yields the following:
\begin{align}
 &\amp_n^\tn{tree}=\frac{\mpn(\Delta_K-\Delta_{K^0})}{2\sqrt{3}F_\pi^4} 
 +\frac{2e^2\mpn}{3\sqrt{3}F_\pi^2}\Bigl(1-\frac{2\epsilon}{\sqrt{3}}\Bigr) \nonumber\\
 &\times \Bigl\{3(2K_3-K_4)-2(K_5+K_6)+2(K_9+K_{10})\Bigr\} \nonumber\\
 &-\frac{1}{Q^22\sqrt{3}F_\pi^2\mpn}\Biggl\{\frac{1}{8F_\pi^2}\biggl[(3\me+\mpn)\nonumber\\
 &\quad \times
\biggl\{ \frac{(\me-5\mpn)(3\me+\mpn)}{\me-\mpn}\times \nonumber\\
&\qquad
\Bigl(\Delta_K\!+\!\Delta_{K^0}\!-\!\Delta_\pi\!+\!\frac{\Delta_{\pi^0}}{2}\!-\!\frac{3}{2}\Delta_\eta\Bigr)
- 4(\me\!-\!\mpn)\Delta_\eta
\nonumber\\
&\quad + \frac{3\me\!-\!5\mpn}{2}(\Delta_K\!-\!\Delta_{K^0}) 
-7\mpn(2\Delta_\pi\!-\!\Delta_{\pi^0}\!-\!\Delta_\eta) \biggr\}
\nonumber\\
&\quad +3\bigl(9M_\eta^4-2\me\mpn+M_{\pi^0}^4\bigr)(\Delta_\pi-\Delta_{\pi^0})
\biggr]\nonumber\\
 &\quad+(3\me+\mpn)\biggl[\frac{3}{8}(\me-\mpn)+\mpn\Delta_F\biggr]\nonumber\\
 &\quad+(3M_\eta^4-9\me\mpn-2M_{\pi^0}^4)\biggl[\frac{\Delta_\tn{GMO}}{2}+\frac{\Delta\mpi}{\me-\mpn}\nonumber\\
 &\qquad \times\!\biggl\{2\Delta_F +\frac{1}{F_\pi^2}\Bigl(\Delta_K-\Delta_\pi
+\frac{\Delta_{K^0}}{2}-\frac{5}{4}\Delta_{\pi^0}+\frac{3}{4}\Delta_\eta\Bigr)\!\biggr\}\nonumber\\
 &\qquad+3e^2\biggl(\frac{\Delta_K-\Delta_\pi}{\me-\mpn}-\frac{1}{16\pi^2}-2(K_{10}+K_{11})\biggr)\biggr]\nonumber\\
 &\quad-e^2\biggl[ (\me-\mpn)(3\me+\mpn)(K_1+K_2) \nonumber\\
 &\qquad -(3\me-5\mpn)(5\me+\mpn)\Bigl(K_3-\frac{K_4}{2}\Bigr) \nonumber\\
 &\qquad +\me(3\me+\mpn)(K_5+K_6) \nonumber\\
 &\qquad -3\me(\me-\mpn)(K_9+K_{10})
\biggr]\Biggr\} ~.
\end{align}

\setcounter{figure}{0}
\setcounter{table}{0}
\section{Loop functions}\label{app:loops}

In this section we give explicit expressions for the relevant loop integrals,
performed in $d$ dimensions, with a renormalization scale $\mu$.

\subsection{Meson loop functions}\label{app:mesonloops}

We define the basic meson loop functions
\begin{align}
 \Delta_a &=\frac{1}{i}\int\frac{d^dk}{(2\pi)^d}\frac{1}{M_a^2-k^2}\;,
 \\
 J_{ab}(s) &=\frac{1}{i}\int\frac{d^dk}{(2\pi)^d}\frac{1}{[M_a^2-k^2][M_b^2-(k-q)^2]}\;,\nonumber
\end{align}
with $s=q^2$.
The tadpole function $\Delta_a$ is given by
\begin{equation}\label{eqn:tadpole}
 \Delta_a=M_a^2\left\{2\lambda+\frac{1}{16\pi^2}\ln\frac{M_a^2}{\mu^2}\right\}\;,
\end{equation}
with the constant $\lambda$ containing the divergent part in $(d-4)$
\begin{equation}
 \lambda=\frac{\mu^{d-4}}{16\pi^2}\left\{\frac{1}{d-4}-\frac{1}{2}(\ln4\pi+\Gamma'(1)+1)\right\}\;,
\end{equation}
where $-\Gamma'(1)=\gamma_E\approx0.5772$ is the Euler--Mascheroni constant. 
The loop function $J_{ab}(s)$ can be split up according to
\begin{align}
 \bar{J}_{ab}(s)&=J_{ab}(s)-J_{ab}(0)\;,\nonumber
 \\
 J_{ab}(0)&=-2\lambda-2k_{ab}\;, \label{eqn:loopJab1}
\end{align}
with the scale-dependent constant $k_{ab}$ and the function $\bar{J}_{ab}(s)$,
\begin{align}
 k_{ab}&=\frac{1}{32\pi^2\Delta_{ab}}\left\{M_a^2\ln\frac{M_a^2}{\mu^2}-M_b^2\ln\frac{M_b^2}{\mu^2}\right\}\;,\nonumber
 \\
 \bar{J}_{ab}(s)&=-\frac{1}{16\pi^2}\Bigg\{
\left(\frac{\Delta_{ab}}{s}-\frac{\Sigma_{ab}}{\Delta_{ab}}\right)\ln\frac{M_a}{M_b}-1\nonumber\\
 +&\frac{\nu}{2s}\bigg[\ln\frac{s-\Sigma_{ab}+\nu}{s-\Sigma_{ab}-\nu}
-2\pi i\,\theta\bigl(s-(M_a\!+\!M_b)^2\bigr)\bigg]\theta(\nu^2)\nonumber\\
 +&\frac{\sqrt{-\nu^2}}{s}\bigg[\arctan\frac{\Delta_{ab}\!+\!s}{\sqrt{-\nu^2}}
-\arctan\frac{\Delta_{ab}\!-\!s}{\sqrt{-\nu^2}}\bigg]\theta(-\nu^2)
 \Bigg\} \,, \label{eqn:loopJab2}
\end{align}
where we have used the abbreviations 
$\Sigma_{ab}=M_a^2+M_b^2$ and $\Delta_{ab}=M_a^2-M_b^2$, 
the usual Heaviside function $\theta(x)$, and 
$\nu^2=\lambda(s,M_a^2,M_b^2)$.

\subsection{Triangle loop function with virtual photon}\label{app:photonloops}

For the triangle diagram depicted in Fig.~\ref{fig:fphotonloops}b) we need the integral 
involving two charged-pion propagators and one photon propagator,
\begin{align}
G(s) &= \\
\frac{1}{i}\int&\frac{d^dk}{(2\pi)^d}\frac{1}{[\mpi-(k+q_a)^2][\mpi-(k-q_b)^2][k^2-m_\gamma^2]} \,, \nonumber
\end{align}
where $s=(q_a+q_b)^2$ and we only consider the on-shell case $q_{a/b}^2=\mpi$. 
$G(s)$ can be written, in the relevant kinematical region $s>4\mpi$ above threshold, as
\begin{align}\label{eqn:triangleloop}
 G(s)&=\frac{1}{32\pi^2s\sigma}\bigg\{4\Li\left(\frac{1-\sigma}{1+\sigma}\right)-2\pi^2\nonumber\\
 &\quad+4\ln\left(\frac{1-\sigma}{1+\sigma}\right)\ln\left(\frac{2\sigma}{1+\sigma}\right)
-\ln^2\left(\frac{1-\sigma}{1+\sigma}\right)\nonumber\\
 &\quad-2 \ln\left(\frac{\sigma^2s}{m_\gamma^2}\right)
 \left[\ln\frac{1-\sigma}{1+\sigma}+i\pi\right]\biggr\} + \order(m_\gamma^2)\;,
\end{align}
with the dilogarithm or Spence's function
\begin{equation}
 \Li(z)=\int_{1}^{z}\frac{\ln t}{1-t}dt\;.
\end{equation}
In this representation the real part of the Coulomb pole, i.e.\ the kinematical divergence at threshold, 
resides solely in the term proportional to $-2\pi^2$.

\setcounter{figure}{0}
\setcounter{table}{0}
\section{Numerical parameters}\label{app:numbers}

For the meson masses we use
$M_{\pi^0}=134.98\mev$, 
$M_{\pi}=139.57\mev$, 
$M_{K^0}=497.61\mev$, 
$M_K=493.68\mev$ and 
$M_\eta=547.85\mev$~\cite{pdg}.
For the meson decay constants, we employ
$F_\pi=92.2\mev$ and $F_K=1.193\,F_\pi$~\cite{pdg},
the latter value relying on Standard Model electroweak couplings~\cite{BernardPassemar}.
The electric charge and the quark mass double ratio are evaluated using
$\alpha = e^2/4\pi = 1/137.036$ and $Q=Q_D=24.2$. 
The photon cutoff energy $\Eg_\tn{cut}$ is set to a typical detector resolution of $10\mev$.

For the only strong low-energy constant not expressed in terms of physical observables,
we use the value $L_3=-3.5\times10^{-3}$ from Ref.~\cite{beglec}. 
As we concentrate on the electromagnetic corrections in this article, 
we do not consider uncertainties for $L_3$.
For the electromagnetic low-energy constants $K_i^r$, we rely on the estimates
in Refs.~\cite{mousslec,ananmousslec,haefelilec} 
(compare Refs.~\cite{bulec,bijpradlec,pinzke} for alternatives)
using the Feynman gauge and given at a scale
$\mu = M_\rho = 0.77\gev$: 
$-K_1^r=K_3^r=2.7\times10^{-3}$, 
$4K_2^r=2K_4^r=K_6^r=2.8\times10^{-3}$, 
$K_7^r=K_8^r=0$, 
$K_{10}^r=4.0\times10^{-3}$, and 
$K_{11}^r=1.3\times10^{-3}$.
Since no number for $K_9^r$ is offered, we go back to an earlier evaluation in Ref.~\cite{bulec}
and use $K_9^r=0$.
The uncertainties in the $K_i^r$ are difficult to assess; 
they are the dominant sources of uncertainties for the electromagnetic corrections.
We adopt the following procedure.  
Since the values of the renormalized LECs at two different scales are related by
\begin{equation}
 K_i^r(\mu_2)=K_i^r(\mu_1)+\frac{\Sigma_i}{16\pi^2}\ln\frac{\mu_1}{\mu_2}\;,
\end{equation}
it appears natural to use correlated errors due to a variation of scale according to 
\begin{equation}
 K_i^r\to K_i^r\pm\frac{\Sigma_i}{16\pi^2}\;. \label{eqn:SigmaError}
\end{equation}
This procedure has the advantage that the resulting error estimate is invariant
under a redefinition of the Lagrangian~\eqref{eqn:l4em}, 
in contrast to a more naive uncorrelated variation of all $K_i^r$ according to
$K_i^r\pm 1/ 16\pi^2$.
However, in some cases (most notably for the normalization of the DKM amplitude for 
$\eta\to\pi^+\pi^-\pi^0$), an accidental cancellation between the various 
$\Sigma_i$ in~\eqref{eqn:SigmaError} leads to an unrealistically small error.
In those cases, we have replaced the $\Sigma_i$ by $1$ in~\eqref{eqn:SigmaError}
and marked the corresponding errors, obtained in a non-standard way,
by an asterisk in Tables~\ref{tab:cdalitzpar}, \ref{tab:decaywidths}, and \ref{tab:Q}.

\end{appendix}




\end{document}